\documentclass[12pt]{article}
\usepackage{amssymb,amsfonts,latexsym}

\newcommand{\si}{\phi}
\newcommand{\sv}{\Phi}
\newcommand{\e}{\eta}
\newcommand{\gs}{{h(\e,\sv)}}
\newcommand{\gst}{{h(\e,\sv,\s)}}
\newcommand{\gr}{{\gb_{\r(\s)}}}
\newcommand{\sgb}{{\mbox{\scriptsize{\gb}}}}
\newcommand{\shb}{{\mbox{\scriptsize{\hb}}}}

\newcommand{\diag}{{\textup{\scriptsize{diag}}}}
\newcommand{\sdiag}{{\textup{\tiny{diag}}}}

\newcommand{\lr}{{\mathcal{L}}}

\renewcommand{\j}{{\mathcal{J}}}

\renewcommand{\theequation}{\arabic{section}.\arabic{equation}}
\renewcommand{\(}{\begin{equation}}
\renewcommand{\)}{end{equation} \vspace{-.05in}\linebreak}

\newcounter{saveeqn}
\newcounter{savealpheqn}

\newcommand{\alpheqn}{\setcounter{saveeqn}{\value{equation}}%
 \stepcounter{saveeqn}\setcounter{equation}{0}%
 \renewcommand{\theequation}{\mbox{\arabic{section}.\arabic{saveeqn}\alph{equation}}}
 \renewcommand{\)}{\end{equation}}}
\def\group#1{\refstepcounter{equation}\setcounter{saveeqn}{\value{equation}}%
 \label{#1}\setcounter{equation}{0}%
 \renewcommand{\theequation}{\mbox{\arabic{section}.\arabic{saveeqn}\alph{equation}}}
 \renewcommand{\)}{\end{equation}}}
\newcommand{\reseteqn}{\setcounter{equation}{\value{saveeqn}}%
 \renewcommand{\theequation}{\arabic{section}.\arabic{equation}}%
 \renewcommand{\)}{\end{equation}}}
\newcounter{alphcount}
\def\getletter#1{\renewcommand{\theequation}{\alph{equation}}%
		 \setcounter{alphcount}{\value{equation}}
		 \refstepcounter{equation}%
                 \label{#1}%
		 \setcounter{equation}{\value{alphcount}}%
                 \renewcommand{\theequation}{\mbox{\arabic{section}.\arabic{saveeqn}\alph{equation}}}}
\def\writeletter#1{\renewcommand{\theequation}{\alph{#1}}%
		 \begin{eqnarray}%
                 \label{#1}%
		 \nonumber\end{eqnarray}\vspace{-.666in}}
\def\getlette2r#1{\newcounter{#1}%
		 \setcounter{#1}{\value{equation}}%
	         \providecommand{\writeletters}{\writeletters\writeletter{#1}}}

\newcommand{\aalpheqn}{\setcounter{saveeqn}{\value{equation}}%
 \stepcounter{saveeqn}\setcounter{equation}{0}%
 \renewcommand{\theequation}{\mbox{\Alph{subsection}.\arabic{saveeqn}\alph{equation}}}
  \renewcommand{\)}{\end{equation}}}
\newcommand{\areseteqn}{\setcounter{equation}{\value{saveeqn}}%
 \renewcommand{\theequation}{\Alph{subsection}.\arabic{equation}}%
 \renewcommand{\)}{\end{equation}}}

\renewcommand{\=}{\hspace{-.03in}=\hspace{-.02in}}
\newcommand{\+}{\hspace{-.03in}+\hspace{-.02in}}
\renewcommand{\thefootnote}{\alph{footnote}}
\renewcommand{\(}{\begin{equation}}
\renewcommand{\)}{\end{equation}}
\newcommand{\ba}{\begin{eqnarray}}
\newcommand{\ea}{\end{eqnarray}}
\renewcommand{\l}{\lambda}

\renewcommand{\a}{\alpha}
\renewcommand{\b}{\beta}
\renewcommand{\r}{\rho}
\newcommand{\sa}{\mathop{\vtop{\ialign{##\crcr
  $\hfil\displaystyle{\longleftrightarrow}\hfil$\crcr\noalign{\kern-1pt\nointerlineskip}
  \hspace{.12in}$^\sigma$\hskip6pt\crcr\noalign{\kern3pt}}}}}
\newcommand{\bp}{\mathop{\vtop{\ialign{##\crcr
  $\hfil\displaystyle{}\hfil$\crcr\noalign{\kern-13pt\nointerlineskip}
  \BIG{(}\hskip0pt\crcr\noalign{\kern3pt}}}}}
\newcommand{\cbp}{\mathop{\vtop{\ialign{##\crcr
  $\hfil\displaystyle{}\hfil$\crcr\noalign{\kern-13pt\nointerlineskip}
  \BIG{)}\hskip0pt\crcr\noalign{\kern3pt}}}}}
\newcommand{\pa}{\mathop{\vtop{\ialign{##\crcr
  $\hfil\displaystyle{\oplus}\hfil$\crcr\noalign{\kern+1pt\nointerlineskip}
  \hspace{.08in}$^{\alpha=0}$\hskip6pt\crcr\noalign{\kern3pt}}}}}
\newcommand{\pan}{\mathop{\vtop{\ialign{##\crcr
  $\hfil\displaystyle{\oplus}\hfil$\crcr\noalign{\kern+2pt\nointerlineskip}
  \hspace{.03in} $^{\alpha}$\hskip6pt\crcr\noalign{\kern3pt}}}}}

\newcommand{\s}{\sigma}

\renewcommand{\sp}{,\hspace{.3in}}
\newcommand{\newsection}{\setcounter{equation}{0}\section}
\newcommand{\p}{^\prime}
\newcommand{\pp}{^{\prime\prime}}

\newcommand{\w}{\omega}

\renewcommand{\lor}{\frac{\l}{\r}}
\newcommand{\loe}{\frac{\l}{\e}}
\newcommand{\lors}{\frac{\l}{\r(\s)}}

\newcommand{\mod}{{\textup{\scriptsize{ mod }}}}
\newcommand{\rank}{{\textup{\scriptsize{rank}}}}
\newcommand{\rrank}{{\textup{rank}}}

\newcommand{\reg}{O((z-w)^0)}
\newcommand{\jh}{\hat{J}}

\newcommand{\hc}{$\hat{J}_{\gst}$}
\newcommand{\tp}{{2\pi i}}
\newcommand{\az}{\frac{A(\zl)}{\zl}}

\newcommand{\taz}{$A(\zl)/\zl$}

\newcommand{\dg}{\textup{dim}\gb}
\renewcommand{\dh}{\textup{dim}\hb}
\newcommand{\arange}{a,b=1,...,\dg}
\newcommand{\appendixa}
 {\renewcommand{\theequation}{\Alph{subsection}.\arabic{equation}}%
  \renewcommand{\thesubsection}%
               {Appendix \Alph{subsection}.\setcounter{equation}{0}}%
  \renewcommand{\alpheqn}{\aalpheqn}%
  \renewcommand{\reseteqn}{\areseteqn}
  \newcounter{savesec}}
\newcommand{\appendices}{\appendix\appendixa}
\def\app#1#2{\renewcommand{\thesubsection}{\Alph{subsection}}%
	\refstepcounter{subsection}%
	\setcounter{subsection}{\value{savesec}}%
	\stepcounter{savesec}\label{#1}%
	\renewcommand{\thesubsection}%
               {Appendix \Alph{subsection}.\setcounter{equation}{0}}%
	\subsection{#2}}

\def\BIG#1{\mbox{\Huge $#1$}}
\def\gb            {\mbox{$\mathfrak g$}}
\def\hb            {\mbox{$\mathfrak h$}}
\def\ssz	   {\mbox{\tiny $\mathbb  Z$}}
\def\sz		   {\mbox{\scriptsize $\mathbb  Z$}}
\def\z		   {\mbox{$\mathbb  Z$}}
\def\zl		   {\mbox{$\mathbb  Z_\l$}}

\mathchardef\endbar="375
\font\fivesans=cmss10 at 4.61pt
\font\sevensans=cmss10 at 6.81pt
\font\tensans=cmss10 at 12pt %added ``at 12pt''
\newfam\sansfam
\textfont\sansfam=\tensans\scriptfont\sansfam=\sevensans\scriptscriptfont
\sansfam=\fivesans

\font\tensans=cmss10 at 14pt 
\newfam\sansfamb
\textfont\sansfamb=\tensans\scriptfont\sansfamb=\sevensans\scriptscriptfont
\sansfamb=\fivesans

\font\tensans=cmss10 at 17pt
\newfam\sansfamc
\textfont\sansfamc=\tensans\scriptfont\sansfamc=\sevensans\scriptscriptfont
\sansfamc=\fivesans

\font\tensans=cmss10 at 10pt
\def\contr#1#2{\mathop{\vtop{\ialign{##\crcr
  $\hfil\displaystyle{#2}\hfil$\crcr\noalign{\kern3pt\nointerlineskip}
  \hspace{.09in}\rule[0in]{.01in}{.1in}\rule[0in]{#1in}{.01in}\rule[0in]{.01in}{.1in}\hskip6pt\crcr\noalign{\kern3pt}}}}}
\def\contrb#1#2#3{\mathop{\vtop{\ialign{##\crcr
  $\hfil\displaystyle{#3}\hfil$\crcr\noalign{\kern3pt\nointerlineskip}
  \hspace{#1in}\rule[0in]{.01in}{.1in}\rule[0in]{#2in}{.01in}\rule[0in]{.01in}{.1in}\hskip6pt\crcr\noalign{\kern3pt}}}}}

\def\hsp#1{\hspace{#1in}}
\def\rf#1{\ref{ref#1}}
\def\comment#1{\hsp{.3}\textup{#1}}

\catcode`\@=11
\def\vereq#1#2{\lower3pt\vbox{\baselineskip1.5pt \lineskip1.5pt
\ialign{$\m@th#1\hfill##\hfil$\crcr#2\crcr\sim\crcr}}}
\catcode`\@=12

\begin{document}
\begin{titlepage}
\begin{center}
December 9, 1999           \hfill UCB-PTH-99/53   \\
                                \hfill LBNL-44692    \\
%                                \hfill hep-th/9904105    \\

\vskip .25in
\def\thefootnote{\fnsymbol{footnote}}

{\large \bf Cyclic Coset Orbifolds \\}

\vskip 0.3in

J. EVSLIN, M. B. HALPERN, and J. E. WANG\footnote{E-Mail: hllywd2@physics.berkeley.edu}

\vskip 0.15in

{\em Department of Physics,
     University of California\\
     Berkeley, California 94720}\\
and\\
{\em Theoretical Physics Group\\
     Ernest Orlando Lawrence Berkeley National Laboratory\\
     University of California,
     Berkeley, California 94720}
        
\end{center}

\vskip .3in

\vfill

\begin{abstract}
We apply the new orbifold duality transformations to discuss the special case of cyclic coset orbifolds in further detail.  We focus in particular on the case of the interacting cyclic coset orbifolds, whose untwisted sectors are $\zl$(permutation)-invariant $g/h$ coset constructions which are not $\l$ copies of coset constructions.  Because $\l$ copies are not involved, the action of $\zl$(permutation) in the interacting cyclic coset orbifolds can be quite intricate.  The stress tensors and ground state conformal weights of all the sectors of a large class of these orbifolds are given explicitly and special emphasis is placed on the twisted $h$ subalgebras which are generated by the twisted (0,0) operators of these orbifolds.  We also discuss the systematics of twisted (0,0) operators in general coset orbifolds.
\end{abstract}

\vfill

\end{titlepage}
\setcounter{footnote}{0}
\renewcommand{\thefootnote}{\alph{footnote}}

%THIS PAGE (PAGE ii) CONTAINS THE LBL DISCLAIMER
%TEXT SHOULD BEGIN ON NEXT PAGE (PAGE 1)
%\renewcommand{\thepage}{\roman{page}}
%\setcounter{page}{1}
%\mbox{ }
%
%\vskip 1in
%
%\begin{center}
%{\bf Disclaimer}
%\end{center}
%
%\vskip .2in
%
%\begin{scriptsize}
%\begin{quotation}
%This document was prepared as an account of work sponsored by the
%United
%States Government. While this document is believed to contain
%correct
% information, neither the United States Government nor any agency
%thereof, nor The Regents of the University of California, nor any
%of their
%employees, makes any warranty, express or implied, or assumes any legal
%liability or responsibility for the accuracy, completeness,
%or usefulness
%of any information, apparatus, product, or process disclosed, or
%represents that its use would not infringe privately owned rights.
%Reference herein
%to any specific commercial products process, or service by
%its trade name,
%trademark, manufacturer, or otherwise, does not necessarily
%constitute or
%imply its endorsement, recommendation, or favoring by the
%United States
%Government or any agency thereof, or The Regents of the
%University of
%California.  The views and opinions of authors expressed herein
%do not
%necessarily state or reflect those of the United States
%Government or any
%agency thereof, or The Regents of the University of California.
%\end{quotation}
%\end{scriptsize}
%
%\vskip 2in
%
%\begin{center}
%\begin{small}
%{\it Lawrence Berkeley National Laboratory is an equal opportunity
%employer.}
%\end{small}
%\end{center}

\pagebreak
\renewcommand{\thepage}{\arabic{page}}
%\tableofcontents
\pagebreak
\newsection{Introduction}
In Ref.~[\rf{us2}] a construction of the stress tensors of all the sectors of all current-algebraic orbifolds
\(
\frac{A(H)}{H}
\)
was given, where $H$ is any finite group and $A(H)$ is any $H$-invariant CFT$^{\rf{4}-\rf{3}}$ constructed on affine Lie algebra$^{\rf{10}-\rf{12}}$.  This technology employs new duality transformations among the sectors of each orbifold, and, in particular, the special case of cyclic permutation orbifolds
\(
\az
\)
was worked out in full as an illustration.  

The untwisted sectors of the permutation orbifolds \taz\ are described by the set of all $\zl$(permutation)-invariant CFT's $A(\zl)$, that is by all the $\zl$(permutation)-invariant affine-Virasoro$^{\rf{1},\rf{2},\rf{3}}$ constructions $L_{I-J}^{ab}$ on affine $g$, where
\(
g=\times_{I=0}^{\l-1}\gb^I\sp\gb^I\cong\gb\sp
\zl\textup{(permutation)}\subset Aut(g). \label{gcopies}
\)
Here $\zl$(permutation) acts by permuting the copies $\{\gb^I\}$ at level $k$ of the affine algebra on simple $\gb$.  In this case the twisted current algebras of the twisted sectors are orbifold affine algebras$^{\rf{9},\rf{5},\rf{us2}}$, and the duality transformations relating the sectors of the orbifolds \taz\ are discrete Fourier transforms.

In this paper, we apply the results of Ref.~[\rf{us2}] for \taz\ to study the set of cyclic coset orbifolds
\group{invarorbs}
\(
\frac{(\mbox{\small$\frac{g}{h(\sz_\l)}$})}{\zl}\subset\az
\)
\(
 h(\zl)\subset g\sp 
\zl(\textup{permutation})\subset Aut(h(\zl)) \label{invarcond}
\)
\reseteqn
in further detail.  These orbifolds, first delineated in Ref.~[\rf{us2}], are the special cases in \taz\ for which the $\zl$(permutation)-invariant CFT $A(\zl)$ is a coset construction$^{\rf{12},\rf{18},\rf{22}}$.  In what follows, an algebra $h(\zl)$ which satisfies (\ref{invarcond}) is called a $\zl$-covariant subalgebra\footnote{Since the stress tensor $T_g$ of $g$ defines a $\zl$(permutation)-invariant CFT, the property (\ref{invarcond}) is a necessary and sufficient condition so that the stress tensors $T_{h(\sz_\l)}$ and $T_{g/h(\sz_\l)}$ also define $\zl$(permutation)-invariant CFT's.} of the ambient algebra $g$.  

Among the cyclic coset orbifolds, the cyclic coset \textit{copy} orbifolds
\(
\frac{(\frac{g}{\times_{I=0}^{\l-1}\shb^I})}{\zl}=\frac{\times_{I=0}^{\l-1}(\gb/\hb)^I}{\zl}\subset\frac{(\mbox{\small$\frac{g}{h(\sz_\l)}$})}{\zl}
\label{copyorbs}
\)
are well-known examples whose stress tensors in the untwisted sectors are sums of $\l$ commuting copies.  In these orbifolds, the action of $\zl$(permutation) consists of cyclic permutations of the coset copies.  The stress tensors of all the twisted sectors of all cyclic copy orbifolds are given in Ref.~[\rf{us2}].  The case of prime $\l$ was given earlier in Refs.~[\rf{9},\rf{5}] and copy orbifolds have also been studied in Refs.~[\rf{39}-\rf{5d}].  As a consequence, our remarks will be minimal for these cases.

We focus instead on a more intricate class of cyclic coset orbifolds called the \textit{interacting} cyclic coset orbifolds$^{\rf{9},\rf{5},\rf{us2}}$ because their stress tensors in the untwisted sectors are \textit{not} sums of $\l$ commuting copies.  As a consequence, the action of $\zl$(permutation) in the interacting coset orbifolds is generally more involved than the action of $\zl$(permutation) in the coset copy orbifolds.  The simplest examples of interacting cyclic coset orbifolds are the cases
\(
\frac{(\mbox{\small$\frac{g}{\sgb_\sdiag}$})}{\zl} \label{kworb}
\)
where $\gb_\diag$ is the diagonal subalgebra of $g$.  The stress tensors and ground state conformal weights of these orbifolds are worked out in Ref.~[\rf{us2}].  As conjectured in Ref.~[\rf{9}], some of the sectors of these orbifolds are Ka\u{c}-Wakimoto coset constructions$^{\rf{13},\rf{14}}$.

Generalizing the simplest examples (\ref{kworb}), we will construct here a large class $\{\gs\}$ of $\zl$-covariant subalgebras 
\(
\{\gs\}\subset \{h(\zl)\}\sp
h(1,\{0\})=\gb_{\diag}
\)
where $\e$ and $\Phi$ are defined below, and give all the stress tensors and ground state conformal weights of the corresponding interacting cyclic coset orbifolds
\(
\frac{(\frac{g}{\gs})}{\zl}\ \subset\ \frac{(\frac{g}{h(\sz_\l)})}{\zl}\ \subset\ \az.
\)
Special emphasis is placed on the twisted $h$ subalgebras (corresponding to each untwisted subalgebra $\gs$) which are generated by the twisted (0,0) operators of these orbifolds.

The systematics of twisted (0,0) operators in general coset orbifolds 
\(
\frac{(g/h)}{H}\ \subset\ \frac{A(H)}{H}\sp
h\ \subset\ g\sp H\ \subset\ Aut(g)\sp
H\ \subset\ Aut(h)
\)
is also discussed, and App.~\ref{copyapp} gives the (0,0) operators of the twisted sectors of the cyclic coset copy orbifolds (\ref{copyorbs}).  The twisted (0,0) operators of coset orbifolds will be important in an action formulation of these theories, where one must learn to gauge these twisted subalgebras.

\newsection{The Untwisted Sectors}
\subsection{Notation}
In what follows, we assume that $\gb$ in (\ref{gcopies}) is simple.  Our notation for the currents of the ambient algebra $g$ is
\alpheqn
\(
J_g(z)=\{J_{a(I)}(z)\}=\{H_{A(I)}(z),E_{\a (I)}(z)\}
\)
\(
a=1,...,\dg\sp
I=0,...,\l-1\sp
A=1,...,\rrank\gb\sp
\a\in\Delta(\gb)
\)
\reseteqn
where the index $I$ labels the copies of $\gb$ and the index $a$ runs over a general basis of $\gb$.  The algebra of affine~$g$ 
\group{affalg}
\(
[H_{A (I)}(m),H_{B (J)}(n)]=\delta_{IJ} k\hsp{.03}m\hsp{.03}\delta_{AB}\delta_{m+n,0}
\)
\(
[H_{A (I)}(m),E_{\a (J)}(n)]=\delta_{IJ}\a_AE_{\a (I)}(m+n)
\)
\(
[E_{\alpha (I)}(m),E_{\beta (J)}(n)]
=\delta_{IJ}\left\{\begin{array}{lll}
N(\alpha,\beta)E_{\alpha+\beta,(I)}(m+n) && \textrm{if}\  (\alpha+\beta)\in\Delta(\gb)\nonumber\\
\alpha\cdot H_{(I)}(m+n)+ k\hsp{.03}m\hsp{.03}\delta_{m+n,0} && \textrm{if}\ (\alpha+\beta)=0\nonumber\\
0 && \textrm{otherwise}\\
\end{array}\right.
\)
\(
A,B=1,...,{\textrm{rank}}\gb\sp
\alpha,\beta\in\Delta(\gb)\sp
I,J=0,...,\l-1\sp
m,n\in\z
\)
\reseteqn
holds in the Cartan-Weyl basis of $\gb$.

The action of the automorphism group $\zl$(permutation) on affine $g$ is
\group{zlact}
\(
J_{a(I)}\ \p=\w_{IJ}(h_\s)J_{a(J)}\sp \w_{IJ}(h_\s)=\delta_{I+\s,J\mod\l}\in\zl(\textup{permutation}) \subset Aut(g) \label{gact}
\)
\(
\s=0,...,\l-1\sp
\r(\s)\in\{1,...,\l\}\sp
\frac{\r(\s)\s}{\l}\in\{0,...,\l-1\}
\label{rhodef}
\)
\reseteqn
where $\r(\s)$ is the order of the element $h_\s\in\zl$(permutation).

\subsection{The subalgebra $\gs\subset g$} 
The subject of this subsection is the construction of a large class of $\zl$-covariant subalgebras $\{h(\zl)\}$.  The members of this class are labeled as $\gs$,  
\(
g\supset\gs\in\{h(\zl)\}\sp
\loe\in\z^+
\)
where $\e$ is any positive integer that divides $\l$, and $\sv$ is constructed for each choice of $\e$ as follows.  On the simple roots $\{\a_i\in\Delta(\gb)\}$ of $\gb$, choose an integer-valued function $\phi(\a_i)$ with values in the range $0$ to $\loe-1$:
\(
\sv\equiv\{\si(\a_i)\}:\hsp{.2} 
\si(\a_i)\in\{0,...,\frac{\l}{\e}-1\}\sp
i=1,...,\hsp{.03}{\textup{rank}}\gb.
\)
Therefore at fixed $\l,\ \e$ and $\gb$ we will construct
\(
N(\l,\e,\gb)=({\mbox{\small{$\loe$}}})^{{}^{\rank\sgb}}
\)
$\zl$-covariant subalgebras $\gs$.  Next, we define a natural extension of $\phi(\a_i)$,
\(
\si(\a)\equiv\sum_{i=1}^{\rank\sgb} n_i(\a)\si(\a_i)
\hsp{.2}\textup{when}\hsp{.2}
\a=\sum_{i=1}^{\rank\sgb} n_i(\a)\a_i\sp
n_i(\a)\in\z
\)  
which is defined on all roots $\a\in\Delta(\gb)$.

Using $\si(\a)$, we may now write the currents $J_{\gs}$ of the subalgebra $h(\eta,\Phi)$
\(
J_{\gs}(z)=\{J_a^R(z;\circ)\}=\{H_A^R(z;\circ),E_\a^R(z;\circ)\}
\)
in terms of the currents $J_{a(I)}$ of the ambient algebra $g$,
\group{gsdefs}
\(
H_A^R(z;\circ)\equiv\sum_{j=0}^{\loe-1}H_{A(\e j+R)}(z)\sp
E_\a^R(z;\circ)\equiv\sum_{j=0}^{\loe-1}e^{\frac{\tp j\si(\a)}{\l/\e}}E_{\a(\e j+R)}(z)
\label{thedefs}
\)
\(
\circ\equiv\left\{\begin{array}{cl}
(\e,\sv) &\textup{for}\ E_\a^R\\
\e &\textup{for}\ H_A^R\\
\end{array}\right. \label{star}
\)
\getletter{perg}
\(
H_{A(I\pm\l)}(z)=H_{A(I)}(z)\sp
E_{\a(I\pm\l)}(z)=E_{\a(I)}(z)\comment{(periodicity for $g$)}
\)
\getletter{pergs}
\(
H_A^{R\pm\e}(z;\circ)= H_A^R(z;\circ)\sp
E_\a^{R\pm\e}(z;\circ)= E_\a^R(z;\circ)\comment{(periodicity for $\gs$)}
\)
\(
\loe\in\z^+\sp
R=0,...,\e-1\sp
\a\in\Delta(\gb)\sp
A=1,...,\textup{rank}\gb.
\)
\reseteqn
The periodicity relations in (\ref{gsdefs} \ref{perg},\ref{pergs}) are conventions.

The currents $J_{\gs}$ have the mode expansions
\(
H_A^R(z;\circ)=\sum_{m\in\sz}H_A^R(m;\circ)z^{-m-1}\sp
E_\a^R(z;\circ)=\sum_{m\in\sz}E_\a^R(m;\circ)z^{-m-1}
\)
and, using the algebra (\ref{affalg}) of affine $g$, one finds the subalgebra $h(\e,\Phi)$
\group{gsalg}
\(
[H_{A}^R(m;\circ),H_{B}^S(n;\circ)]=\delta^{RS}\loe k\hsp{.03}m\hsp{.03}\delta_{AB}\delta_{m+n,0}
\)
\(
[H_{A}^R(m;\circ),E_{\a}^S(n;\circ)]=\delta^{RS}\a_AE_{\a}^R(m+n;\circ)
\)
\(
[E_{\alpha}^R(m;\circ),E_{\beta}^S(n;\circ)]\hspace{2.5in}
\)
\[\hspace{.8in}
=\delta^{RS}\left\{\begin{array}{lll}
N(\alpha,\beta)E_{\alpha+\beta}^R(m+n;\circ) && \textrm{if}\  (\alpha+\beta)\in\Delta(\gb)\\
\alpha\cdot H^R(m+n;\circ)+\loe k\hsp{.03}m\hsp{.03}\delta_{m+n,0} && \textrm{if}\ (\alpha+\beta)=0\\
0 && \textrm{otherwise}\\
\end{array}\right.
\]
\(
A,B=1,...,{\textrm{rank}}\gb\sp
\alpha,\beta\in\Delta(\gb)\sp
R,S=0,...,\e-1\sp
m,n\in\z.
\)
\reseteqn
This algebra consists of $\e$ commuting copies of affine $\gb$ at level $\loe k$, where the copies are labeled by the indices $R$ and $S$.

The algebra (\ref{gsalg}) of $\gs$ has no explicit $\si(\a)$ dependence, but the induced action of the $\zl$ automorphism (\ref{zlact}) on the currents $J_{\gs}$,
\(
H_A^R(m;\circ)\ \p=H_A^{R+\s}(m;\circ)\sp
E_\a^R(m;\circ)\ \p=e^{-\frac{\tp\lfloor\frac{R+\s}{\e}\rfloor \si(\a)}{\l/\e}}E_\a^{R+\s}(m;\circ) \label{bigomega}
\)
is different for each $\gs$ (see App.~\ref{lameapp}).  In (\ref{bigomega}), the symbol $\lfloor x\rfloor$ is the greatest integer less than or equal to $x$.  A short calculation shows that $E_\a^R\ \p$ and $H_\a^R\ \p$ in (\ref{bigomega}) also satisfy the algebra (\ref{gsalg}), which means that $\zl$(permutation) is in Aut$(\gs)$.  It follows that each $h(\e,\sv)$ is a $\zl$-covariant subalgebra of $g$.

As an example of these subalgebras\footnote{The trivial case $h(\l,\Phi)=g$ is not useful in the $g/h$ coset constructions below.}, consider the subalgebra $h(1,\{0\})$ for all $\gb$ and $\l$, where $\{0\}$ means $\si(\a_i)=0$, $\forall\a_i$.  For this choice, one finds that
\alpheqn
\(
H_A^{R=0}(z;\circ)=\sum_{I=0}^{\l-1}H_{A(I)}(z)\sp
E_\a^{R=0}(z;\circ)=\sum_{I=0}^{\l-1}E_{\a(I)}(z)
\)
\(
h(\e=1;\sv=\{0\})=\gb_{\mbox{\scriptsize{diag}}}
\)
\reseteqn
so that the $\zl$-covariant subalgebra is the diagonal subalgebra of $g$.  This is the $h$ subalgebra of the simple case in (\ref{kworb}).  

As another example consider the ambient algebra 
\(
g=\times_{I=0}^2\ \gb^I\sp
\gb^I\cong\gb=SU(2)
\)
with simple root $\a$.  This has a $\zl$-covariant subalgebra $h(\e=1,\{\si(\a)=1\})$ whose currents
\group{newex}
\(
H^{R=0}(z;\circ)=H_{(0)}(z)+H_{(1)}(z)+H_{(2)}(z)\sp
\)
\(
E_{\pm\a}^{R=0}(z;\circ)=E_{\pm\a(0)}(z)+e^{\pm\frac{\tp}{3}}E_{\pm\a(1)}(z)+e^{\mp\frac{\tp}{3}}E_{\pm\a(2)}(z)
\)
\reseteqn
generate an affine $SU(2)$ at level $3k$.  Here $H_{(I)}$ and $E_{\pm\a(I)}$, $I=0,1,2$ are the Cartan generator and root operators of $SU(2)^I$.

\subsection{Affine-Sugawara and coset constructions}

The affine-Sugawara construction$^{\rf{12},\rf{18},\rf{19}-\rf{21}}$ on $g$ is
\group{ascon}
\(
T_g=\frac{\e^{ab}}{2k+Q}\sum_{I=0}^{\l-1}:J_{a(I)}J_{b(I)}:
\)
\(
c_g=\frac{\l x \dg}{x+\tilde{h}}\sp
x\equiv\frac{2k}{\psi^2}\in\z^+\sp
\tilde{h}\equiv\frac{Q}{\psi^2}
\)
\(
a,b=1,...,\dg
\)
\reseteqn
where $x$, $\tilde{h}$, $\psi$ and $Q$ are respectively the invariant level, dual Coxeter number, highest root and quadratic Casimir of $\gb$.  In the text below, including the twisted sectors of the interacting coset orbifolds, these quantities always refer to $\gb$ and affine $\gb$.  

The reader should bear in mind that, here and throughout this paper,\ $:\hsp{-.05}(\cdot)\hsp{-.05}:$ means OPE normal ordering$^{\rf{wick},\rf{5},\rf{us2}}$.

The affine-Sugawara construction on the subalgebra $\gs$ is
\alpheqn
\(
T_\gs=\frac{1}{2\loe k+Q}\sum_{R=0}^{\e-1}(\sum_{A=1}^{\rank\sgb}:H_A^R(\circ)H_A^R(\circ):+\sum_{\a\in\Delta(\sgb)}:E_\a^R(\circ)E_{-\a}^R(\circ):) \label{ustress}
\)
\(
c_\gs=\frac{\l x \dg}{\loe x+\tilde{h}}
\)
\reseteqn
because the algebra $\gs$ is embedded at level $\loe k$ in affine $g$.  In what follows we will suppress the explicit $A$ and $\a$ summations above.  

Another form of this stress tensor is
\group{invarform}
\(
T_\gs=\sum_{J,L=0}^{\l-1}[L_{J-L}:H_{A(J)}H_{A(L)}:+L_{J-L}^\a:E_{\a(J)}E_{-\a(L)}:] 
\)
\(
L_{J-L}\equiv\frac{1}{2\loe k+Q}\delta_{J-L,0\mod\e}\sp
L_{J-L}^{\a}\equiv
\frac{1}{2\loe k+Q} e^{\frac{\tp (J-L)\si(\a)}{\l}}\delta_{J-L,0\mod\e}
\label{invarl}
\)
\reseteqn
in terms of the ambient currents $J_g$.  The form in (\ref{invarform}) shows that all the subalgebra constructions $T_\gs$ are $\zl$(permutation)-invariant CFTs because the inverse inertia tensors $L$ depend only on the combination $J-L$ (see Ref.~[\rf{us2}]).

Finally we may use K-conjugation$^{\rf{12},\rf{18},\rf{22},\rf{23},\rf{1}}$ to obtain the coset constructions$^{\rf{12},\rf{18},\rf{22}}$
\alpheqn
\(
T_{g/\gs}=T_g-T_{\gs} \label{untwistcoset}
\)
\(
c_{g/\gs}=c_g-c_{\gs}=\l x\dg(\frac{1}{x+\tilde{h}}-\frac{1}{\loe x+\tilde{h}})
\)
\reseteqn
which describe the untwisted sectors of this class of interacting cyclic coset orbifolds.

\newsection{Twisted Currents}
\subsection{The ambient twisted algebra $\gr$} \label{twistsec}
In Ref.~[\rf{us2}], a general prescription is given to obtain the twisted ambient algebra of each sector $\s$ of any orbifold $A(H)/H$.  One begins with the action of the automorphisms on the ambient currents $J_g$
\(
J_g\ \p=\w(h_\s)J_g\sp
h_\s\in H\subset Aut(g).
\)
The matrix $\w(h_\s)$ defines an eigenvalue problem whose eigenvectors $U_g(\s)$ generate the $g$ eigencurrents
\(
\j_g(\s)\sim U_g(\s)J_g
\)
with diagonal response to the automorphism $\w(h_\s)$.  The twisted $g$ currents $\hat{J}_g(\s)$, which are the ambient currents of each twisted sector $\s$,
\(
\j_g(\s)\rightarrow\hat{J}_g(\s) \label{garrow}
\)
then satisfy the same OPE's as the eigencurrents.  The diagonal monodromies of $\hat{J}_g(\s)$ are controlled by the eigenvalues of $\w(h_\s)$.

In the case$^{\rf{us2}}$ of the cyclic permutation orbifolds \taz, the matrices $\w(h_\s)$ are given in Eq.~(\ref{zlact}), and one obtains the twisted $g$ currents $\hat{J}_{\sgb_{\r(\s)}}$ of sector $\s$, which satisfy the orbifold affine algebra $\gb_{\r(\s)}$:
\group{oaa}
\(
\hat{J}_{\sgb_{\r(\s)}}(z)=\{\hat{J}_{a(j)}^{(r)}(z)\}=\{\hat{H}_{A(j)}^{(r)}(z),\hat{E}_{\a (j)}^{(r)}(z)\}
\)
\(
\jh_{a(j)}^{(r)}(z)\jh_{b(l)}^{(s)}(w) \hspace{-.03in} = \hspace{-.03in} \delta_{j l}[\frac{\r(\sigma) k\e_{ab}\delta_{r+s,0\mod\r(\sigma)}}{(z-w)^2}+\frac{if_{ab}^{\ \ c}\jh_{cj}^{(r+s)}(w)}{(z-w)}]+\reg \label{TCOPE}
\)
\(
\hat{J}_{a(j)}^{(r)}(z) = \sum_{m \in \sz} \hat{J}_{a(j)}^{(r)}(m+\frac{r}{\r(\sigma)}) z^{-m-1-\frac{r}{\r(\sigma)}}\sp
\hat{J}_{a(j)}^{(r)}(ze^{\tp})=e^{-\frac{\tp r}{\r(\s)}}\hat{J}_{a(j)}^{(r)}(z)
\)
\(
\hat{J}_{a(j)}^{(r)}(m+\frac{r}{\r(\s)})|0\rangle_\s=0\textup{\ when\ }m+\frac{r}{\r(\s)}\geq 0
\)
\(
\gb_{\r(\s)}\equiv\times_{j=0}^{\lors-1}\gb_{\r(\s)}^j\sp
\s=0,...,\l-1 \label{grho}
\)
\(
\arange\sp
r,s=0,...,\r(\sigma)-1\sp
j,l=0,...,\frac{\l}{\r(\sigma)}-1.
\)
\reseteqn
Here $j,\ l$ label copies $\gb_{\r(\s)}^j$ of an orbifold affine algebra on simple $\gb$, with twist classes labeled by $r,\ s$.  The quantity $\r(\s)$ is the order of each copy and also the order of the automorphism $h_\s\in\zl$.  The orbifold affine level of each $\gb_{\r(\s)}^j$ is 
\(
\hat{k}(\s)\equiv\r(\s)k
\)
where $k$ is the level of the affine algebra (\ref{affalg}).  In what follows, $\gb_{\r(\s)}$ is called the ambient orbifold affine algebra of sector $\s$.

\subsection{General twisted subalgebras} \label{twistsubsec}
A central problem which arises in general coset orbifolds
\(
\frac{(g/h)}{H}\subset\frac{A(H)}{H}\sp
h\subset g\sp H\subset Aut(g)\sp H\subset Aut(h)
\)
is to find the twisted $h$ currents with diagonal monodromy.  The analysis of Ref.~[\rf{us2}] for the ambient algebra $g$ can also be applied, mutatis mutandis, to the $H$-covariant subalgebra $h$, with attention to the embedding of affine $h$ in affine $g$
\(
J_h=MJ_g
\)
where $M$ is the embedding matrix in the untwisted sector.

The induced action of the automorphism group $H$ on the currents $J_h$ has the form
\(
J_h{}\p=\Omega(\s)J_h
\)
and the matrix $\Omega(\s)$ defines an induced eigenvalue problem whose eigenvectors $U_h(\s)$ generate the $h$ eigencurrents
\(
\j_h(\s)\sim U_h(\s)J_h=U_h(\s)MJ_g=U_h(\s)MU_g^{\dagger}(\s)\j_g(\s)
\label{heigen}
\)
whose response to $\Omega(\s)$ is diagonal.  The twisted $h$ currents $\hat{J}_h$, 
\(
\j_h(\s)\rightarrow\hat{J}_h(\s) \label{harrow}
\)
satisfy the OPE's of the $h$ eigencurrents, with diagonal monodromies controlled by the eigenvalues of $\Omega(\s)$.  Moreover, with Eqs.~(\ref{garrow}), (\ref{heigen}) and (\ref{harrow}), we see that $\hat{J}_h$ is embedded in the ambient twisted algebra as
\(
\hat{J}_h(\s)\sim \hat{M}(\s)\hat{J}_g(\s)\sp
\hat{M}(\s)\equiv U_h(\s)MU_g^{\dagger}(\s)
\)
where $\hat{M}(\s)$ is the embedding matrix of sector $\s$.

The general twisted $h$ currents $\jh_h$ are further discussed in Subsec.~\ref{zerosec}, and Appendix~\ref{copyapp} gives an application of this procedure to the simple case of the cyclic copy orbifolds.

\subsection{The twisted subalgebra $\gst\subset\gb_{\r(\s)}$}
For our class of interacting cyclic coset orbifolds
\(
\frac{(\frac{g}{\gs})}{\zl}\ \subset\ \frac{(\frac{g}{h(\sz_\l)})}{\zl}\ \subset\ \az
\)
the embedding matrix $M$ of the subalgebra $\gs$ is specified in Eq.~(\ref{thedefs}).  In the discussion below, we will use the notation
\alpheqn
\(
\hat{J}_{\gst}(z)=\{\hat{J}_a^{R,(r)}(z;\circ;\s)\}=\{\hat{H}_A^{R,(r)}(z;\circ;\s),\hat{E}_\a^{R,(r)}(z;\circ;\s)\}
\)
\(
\gst\subset\gb_{\r(\s)}\sp
h(\e,\sv,0)=h(\e,\sv)\sp
\gb_{\r(0)=1}=g
\)
\reseteqn
for the twisted $h$ currents of sector $\s$, which generate the twisted subalgebras $\gst$.

Starting from (\ref{gsdefs}) and (\ref{bigomega}), we obtain the explicit form of the twisted currents \hc,
\group{hopes}
\getletter{hdef}
\(
\hat{H}_A^{R,(r)}(z;\circ;\s)=\sum_{j=0}^{\frac{\l}{\r(\s)\mu}-1}e^{\frac{\tp jrN(\s)P}{\e/\mu}}\hat{H}_{A(\mu j+R)}^{(\frac{\r(\s)\mu}{\e}r)}(z)
\)
\getletter{edef}
\(
\hat{E}_\a^{R,(r)}(z;\circ;\s)=\sum_{j=0}^{\frac{\l}{\r(\s)\mu}-1}e^{\frac{\tp j(N(\s)r-\si(\a))P}{\e/\mu}}e^{\frac{\tp j\si(\a)}{\l/\mu}}\hat{E}_{\a(\mu j+R)}^{(\frac{\r(\s)\s}{\l}\si(\a)+\frac{\r(\s)\mu}{\e}r)}(z)
\)
\(
\mu\equiv\mu(\e,\s)=\textup{gcd}(\e,\lors)\sp
\frac{\e}{\mu(\e,\s)}\in\z^+\sp
\frac{\r(\s)\mu(\e,\s)}{\e}\in\z^+
\label{mudef}
\)
\(
P\equiv P(\e,\s)\sp
P(\e,\s)\frac{\l}{\r(\s)\mu}=1\textup{\ mod\ }\frac{\e}{\mu}\sp
P(\e,\s)\in\{1,...,\frac{1}{\eta/\mu}\}
\)
\(
R,S=0,...,\mu-1\sp
r,s=0,...,\frac{\e}{\mu}-1\sp
\s=0,...,\l-1
\)
\reseteqn
in terms of the currents $J_{\sgb_{\r(\s)}}$ of the ambient orbifold affine algebra (\ref{oaa}), where $z$=gcd($x,y$) is the largest integer such that $x/z,\ y/z\in\z$.  The result (\ref{hopes}) summarizes the embedding matrix of sector $\s$.  

The OPE's of the twisted currents \hc\ follow from those of $\gb_{\r(\s)}$ in (\ref{oaa}):
\group{realhopes}
\(
\hat{H}_A^{R,(r)}(z;\circ;\s)\hat{H}_B^{S,(s)}(w;\circ;\s)=\delta_{RS}\frac{\frac{\l k}{\mu}\delta_{AB}\delta_{r+s,0\mod\frac{\e}{\mu}}}{(z-w)^2}+\reg
\)
\(
\hat{H}_A^{R,(r)}(z;\circ;\s)\hat{E}_{\a}^{S,(s)}(w;\circ;\s)=\delta_{RS}\frac{\a_A\hat{E}_\a^{R,(r+s)}(w;\circ;\s)}{z-w}+\reg
\)
\(
\hat{E}_{\a}^{R,(r)}(z;\circ;\s)\hat{E}_{\b}^{S,(s)}(w;\circ;\s)\hspace{3.5in}
\)
\[\hspace{.8in}
=\delta_{RS}\left\{\begin{array}{lll}
\frac{N(\alpha,\beta)\hat{E}_{\alpha+\beta}^{R,(r+s)}(w;\circ;\s)}{z-w}+\reg && \textrm{if}\  (\alpha+\beta)\in\Delta(\gb)\\
\frac{\frac{\l k}{\mu}\delta_{r+s,0\mod\frac{\e}{\mu}}}{(z-w)^2}+\frac{\alpha\cdot \hat{H}^{R,(r+s)}(w;\circ;\s)}{z-w}+\reg && \textrm{if}\ (\alpha+\beta)=0\\
\reg && \textrm{otherwise}\\
\end{array}\right.
\]
\(
R,S=0,...,\mu-1\sp
r,s=0,...,\frac{\e}{\mu}-1. \label{hoperange}
\)
\reseteqn
These OPE's are isomorphic to the OPE's of $\mu$ copies of order $\e/\mu$ orbifold affine algebra taken at orbifold affine level $\l k/\mu$, where the copies are labeled by the indices $R$ and $S$.  The number of copies and the order can be read from (\ref{hoperange}), and the order $\e/\mu$ also appears in the central terms of (\ref{realhopes}).  As a check on these quantities, recall that
\alpheqn
\(
\textup{(orbifold affine level) = (order)$\cdot$(affine level)} 
\label{tlevel}
\)
\(
(\frac{\l}{\mu}k)=(\frac{\e}{\mu})\cdot(\loe k)
\)
\reseteqn
where (\ref{tlevel}) is a general property of orbifold affine algebra and $\loe k$ is the level of the untwisted algebra $\gs$.

The system (\ref{realhopes}) is distinguished however from an orbifold affine algebra by the monodromies
\group{mono}
\(
\hat{H}_A^{R,(r)}(ze^\tp;\circ;\s)=e^{-\tp\frac{r}{\e/\mu}}\hat{H}_{A}^{R,(r)}(z;\circ;\s)
\)
\(
\hat{E}_{\a}^{R,(r)}(ze^\tp;\circ;\s)=e^{-\tp(\frac{r}{\e/\mu}+\frac{\s\si(\a)}{\l})}\hat{E}_{\a}^{R,(r)}(z;\circ;\s)
\)
\reseteqn
of the twisted $h$ currents \hc, which also follow from (\ref{hopes}).  To see the consequences of this difference, we use the mode expansions
\group{modeex}
\(
\hat{H}_A^{R,(r)}(z;\circ;\s)=\sum_{m\in\sz}z^{-m-1-\frac{r}{\e/\mu}}\hat{H}_{A}^{R,(r)}(m+\frac{r}{\e/\mu})
\)
\(
\hat{E}_{\a}^{R,(r)}(z;\circ;\s)=\sum_{m\in\sz}z^{-m-1-\frac{r}{\e/\mu}-\frac{\s\si(\a)}{\l}}\hat{E}_{\a}^{R,(r)}(m+\frac{r}{\e/\mu}+\frac{\s\si(\a)}{\l})
\)
\reseteqn
which follow from the monodromies in Eq.~(\ref{mono}).  Then, (\ref{realhopes}) and (\ref{modeex}) give the twisted subalgebra $\gst\subset\gb_{\r(\s)}$:
\group{halg}
\(
[\hat{H}_{A}^{R,(r)}(m+\frac{r}{\e/\mu}),\hat{H}_{B}^{S,(s)}(n+\frac{s}{\e/\mu})]=\delta_{AB}\delta_{RS}\frac{\l k}{\mu}(m+\frac{r}{\e/\mu})\delta_{m+n+\frac{r+s}{\e/\mu},0}
\)
\(
[\hat{H}_{A}^{R,(r)}(m+\frac{r}{\e/\mu}),\hat{E}_{\a}^{S,(s)}(n+\frac{s}{\e/\mu}+\frac{\s \si(\a)}{\l})]\hsp{2}
\)
\[
\hsp{2}=\delta_{RS}\a_A\hat{E}_{\a}^{R,(r+s)}(m+n+\frac{r+s}{\e/\mu}+\frac{\s\si(\a)}{\l})
\]
\(
[\hat{E}_{\a}^{R,(r)}(m+\frac{r}{\e/\mu}+\frac{\s \si(\a)}{\l}),\hat{E}_{\b}^{S,(s)}(n+\frac{s}{\e/\mu}+\frac{\s \si(\b)}{\l})]\hspace{2.0in}
\)
\[\hspace{.05in}
=\delta_{RS}\left\{\begin{array}{lll}
N(\alpha,\beta)\hat{E}_{\alpha+\beta}^{R,(r+s)}(m\+n\+\frac{r+s}{\e/\mu}\+\frac{\s\si(\a+\b)}{\l}) && \textrm{if}\  \alpha\+\beta\in\Delta(\gb)\\
\alpha\cdot \hat{H}^{R,(r+s)}(m\+n\+\frac{r+s}{\e/\mu})\+\frac{\l k}{\mu}(m\+\frac{r}{\e/\mu}\+\frac{\s\si(\a)}{\l}) \delta_{m+n+\frac{r+s}{\e/\mu},0} && \textrm{if}\ \alpha\+\beta=0\\
0 && \textrm{otherwise}\\
\end{array}\right.
\]
\(
R,S=0,...,\mu-1\sp
r,s=0,...,\frac{\e}{\mu}-1\sp
\s=0,...,\l-1.
\)
\reseteqn
As we will discuss in the following section, algebras of the form (\ref{halg}) are known in the literature.

\subsection{Doubly-twisted current algebras}
To identify the twisted $h$ subalgebras (\ref{halg}), we first introduce the fundamental weights $\{\l_i\}$ of $\gb$
\(
\l_i\cdot\a_j=\delta_{ij}\frac{\a_i^2}{2}\sp
i,j=1,...,\rrank\gb
\)
and the vector $d$,
\alpheqn
\(
d^A\equiv -\frac{2\e\s}{\mu\l}\sum_{i=1}^{r}\frac{\phi(\a_i)\l_i^A}{\a_i^2}\sp
A=1,...,\rrank\gb
\)
\(
\phi(\a)=-\frac{\mu\l}{\e\s}d\cdot\a. \label{ddota}
\)
\reseteqn
Then, using (\ref{ddota}) we can rewrite the induced action (\ref{bigomega})
\(
H_A^R(m)\ \p=H_A^{R+\s}(m)\sp
E_\a^R(m)\ \p=e^{\frac{\tp\lfloor\frac{R+\s}{\e}\rfloor \mu d\cdot\a}{\s}}E_\a^{R+\s}(m)
\label{dtwistomega}
\)
of $\zl$(permutation) on the subalgebra.  The twisted subalgebra $\gst$ can also be rewritten in this notation as
\group{dtwist}
\(
[\hat{H}_{A}^{R,(r)}(m+\frac{r}{\e/\mu}),\hat{H}_{B}^{S,(s)}(n+\frac{s}{\e/\mu})]=\delta_{AB}\delta_{RS}\frac{\l k}{\mu}(m+\frac{r}{\e/\mu})\delta_{m+n+\frac{r+s}{\e/\mu},0}
\)
\(
[\hat{H}_{A}^{R,(r)}(m+\frac{r}{\e/\mu}),\hat{E}_{\a}^{S,(s)}(n+\frac{s-d\cdot\a}{\e/\mu})]\hsp{2}
\)
\[
\hsp{2}=\delta_{RS}\a_A\hat{E}_{\a}^{R,(r+s)}(m+n+\frac{r+s-d\cdot\a}{\e/\mu})
\]
\(
[\hat{E}_{\a}^{R,(r)}(m+\frac{r-d\cdot\a}{\e/\mu}),\hat{E}_{\b}^{S,(s)}(n+\frac{s-d\cdot\b}{\e/\mu})]\hspace{2.0in}
\)
\[\hspace{.05in}
=\delta_{RS}\left\{\begin{array}{lll}
N(\alpha,\beta)\hat{E}_{\alpha+\beta}^{R,(r+s)}(m\+n\+\frac{r+s-d\cdot(\a+\b)}{\e/\mu}) && \textrm{if}\  \alpha\+\beta\in\Delta(\gb)\\
\alpha\cdot \hat{H}^{R,(r+s)}(m\+n\+\frac{r+s}{\e/\mu})\+\frac{\l k}{\mu}(m\+\frac{r-d\cdot\a}{\e/\mu}) \delta_{m+n+\frac{r+s}{\e/\mu},0} && \textrm{if}\ \alpha\+\beta=0\\
0 && \textrm{otherwise}\\
\end{array}\right.
\]
\(
R,S=0,...,\mu-1\sp
r,s=0,...,\frac{\e}{\mu}-1\sp
\s=0,...,\l-1.
\)
\reseteqn
This form of $\gst$ is recognized as $\mu=\mu(\e,\s)$ commuting copies of a \textit{doubly-twisted} current algebra$^{\rf{5}}$.  

The doubly-twisted current algebras correspond to affine algebras which have been simultaneously twisted both by outer automorphisms $\zl$\ (permutations) and inner automorphisms ($d\neq 0$) of $\gb$.  Equivalently, doubly-twisted current algebras are inner-automorphically twisted orbifold affine algebras.  The doubly-twisted algebras have an order (in this case $\e/\mu$) and a level (in this case $\l k/\mu$), which are the order and level of the orbifold affine algebra$^{\rf{9},\rf{5},\rf{us2}}$ before the inner-automorphic twist.  

The origin of these algebras in this problem can be understood from Eq. (\ref{dtwistomega}), which shows that the induced automorphism $\Omega$ on $J_h$ is a combination of the permutation automorphism $(R\rightarrow R+\s)$ and an inner automorphism (the phases).  In turn, the inner automorphisms can be traced to the form (\ref{thedefs}) of the root operators of the untwisted subalgebra $\gs$
\alpheqn
\(
E_\a^R(z,\circ)=\sum_{j=0}^{\loe-1}E_{\a(\e j+R)}(z)\pp
\)
\(
E_{\a(\e j+R)}{}\pp\equiv
e^{\frac{\tp j\mu d\cdot\a}{\s}}E_{\a(\e j+R)}
= \xi(\e j+R)E_{\a(\e j+R)}\xi(\e j+R)^{-1}\)
\(
\xi(\e j+R)\equiv e^{\tp j\frac{\mu}{\s}d\cdot H_{(\e j+R)}(0)}
\)
\reseteqn
where $E_{\a(\e j+R)}{}\pp$ is inner-automorphically equivalent to $E_{\a(\e j+R)}$.

In special cases, the doubly-twisted subalgebras reduce to singly-twisted subalgebras: When the inner-automorphic vector $d$ vanishes
\(
d=0\ \Leftrightarrow\ \Phi=\{0\}
\)
then the subalgebras $h(\eta,0,\s)$ are orbifold affine algebras (see also Subsec.~\ref{firstex}).  When $d\neq 0$, $\e=1$ (and hence $\mu=1$) the twisted subalgebras are inner-automorphically twisted$^{\rf{47}-\rf{45}}$ affine Lie algebras on simple $\gb$ (see also Subsec.~\ref{fourthex}).  When $d\neq 0$ and
\(
\frac{\l}{\r(\s)\e}\in\z\sp
\mu=\e\neq 1
\)
the twisted subalgebras consist of $\e$ copies of inner-automorphically twisted affine $\gb$ (see also Subsec.~\ref{fifthex}).

We emphasize that the doubly-twisted algebras (\ref{dtwist}) are subalgebras of orbifold affine algebra.  The explicit form of the embedding 
\alpheqn
\(
\hat{H}_A^{R,(r)}(m+\frac{r}{\e/\mu})=\sum_{j=0}^{\frac{\l}{\r(\s)\mu}-1}e^{\frac{\tp jrN(\s)P}{\e/\mu}}\hat{H}_{A(\mu j+R)}^{(\frac{\r(\s)\mu}{\e}r)}(m+\frac{\r(\s)\mu r/\e}{\r(\s)})
\)
\ba
&&\hat{E}_\a^{R,(r)}(m+\frac{r-d\cdot\a}{\e/\mu})\\
&&\hsp{.4}=\sum_{j=0}^{\frac{\l}{\r(\s)\mu}-1}e^{\frac{\tp j(N(\s)r-\si(\a))P}{\e/\mu}}e^{\frac{\tp j\si(\a)}{\l/\mu}}\hat{E}_{\a(\mu j+R)}^{(\frac{\r(\s)\mu}{\e}r+\frac{\r(\s)\s}{\l}\phi(\a))}(m+\frac{\frac{\r(\s)\mu}{\e}r+\frac{\r(\s)\s}{\l}\phi(\a)}{\r(\s)})\nonumber
\ea
\(
\frac{r-d\cdot\a}{\e/\mu}=\frac{\frac{\r(\s)\mu}{\e}r+\frac{\r(\s)\s}{\l}\phi(\a)}{\r(\s)}\sp
\frac{\r(\s)\mu}{\e}r+\frac{\r(\s)\s}{\l}\phi(\a)\in\z^+
\)
\reseteqn
follows from the discussion above.  It has been further observed in Ref.~[\rf{9}] that outer-automorphically twisted affine Lie algebras$^{\rf{46}}$ also occur as subalgebras of orbifold affine algebras.  So the orbifold affine algebras contain (as subalgebras) examples of all standard twisted current algebras.

It may also be possible to find interacting cyclic coset orbifolds whose twisted $h$ subalgebras are ``triply-twisted'' current algebras, which are twisted as well by outer automorphisms of simple $\gb$.

\newsection{Stress tensors of the twisted sectors}
\subsection{Systematics}
The $\zl$-invariant CFT's $A(\zl)$ are described by the stress tensors
\alpheqn
\(
T=L_{I-J}^{ab}:J_{a(I)}J_{b(J)}:\sp\s=0
\)
\(
L^{ab}_K=L^{ba}_{-K}=L^{ab}_{K\pm\l}\sp K=0,...,\l-1 \label{dualsym}
\)
\reseteqn
where $L_{I-J}^{ab}$ is any solution of the Virasoro master equation$^{\rf{1},\rf{2},\rf{3}}$ (VME) with cyclic permutation symmetry.  The explicit form of this consistent subansatz of the VME is given in Ref.~[\rf{us2}].  Then, the duality transformations of Ref.~[\rf{us2}] give the stress tensor for each twisted sector $\s$ of all cyclic permutation orbifolds \taz\ in terms of the twisted currents $\hat{J}_{a(j)}^{(r)}$ of the ambient orbifold affine algebra $\gb_{\r(\s)}$ in Eq.~(\ref{oaa}):
\alpheqn
\(
\hat{T}_\s (z)=\sum_{r=0}^{\r(\sigma)-1} \ \sum_{j,l=0}^{\frac{\l}{\r(\sigma)}-1} \lr_r^{a(j)b(l)}(\s):\jh_{a(j)}^{(r)}(z)\jh_{b(l)}^{(-r)}(z): \sp \s=1,...,\l-1 
\label{twistedstress}
\)
\(
\lr_{r}^{a(j)b(l)}(\s)=\frac{1}{\r(\sigma)}\sum_{s=0}^{\r(\sigma)-1} e^{-\frac{\tp N(\s)rs}{\r(\sigma)}} L^{ab}_{\frac{\l}{\r(\sigma)} s+j-l}. \label{themap}
\)
\reseteqn
The relation in (\ref{themap}) is the duality transformation from the inverse inertia tensor $L$ of the untwisted sector to the inverse inertia tensor $\lr$ of the twisted sector $\s$.  The integers $N(\s)$ and the ground state conformal weights $\hat{\Delta}_0(\s)$
\alpheqn
\(
\hat{T}_\s(z)=\sum_mL_\s(m)z^{-m-2}\sp
L_\s(m\geq 0)|0\rangle_\s=\delta_{m,0}\hat{\Delta}_0(\s)|0\rangle_\s
\)
\ba
\hat{\Delta}_0(\s)&=&\sum_{j=0}^{\lors-1}\r(\s) k \eta_{ab}\sum_{r=0}^{\r(\s)-1}
\lr_r^{a(j) b(j)} (\sigma) \ \frac{r(\r(\s)-r)}{2\r^2(\s)} \label{cwlr} \\
&=&\frac{\l k\eta_{ab}}{4\r^2(\s)}(\frac{\r^2(\s)-1}{3}L_0^{ab}-{\displaystyle{\sum_{s=1}^{\r(\s)-1}}} csc^2(\frac{\pi N(\s)s}{\r(\s)})\hsp{.02} L^{ab}_{\frac{\l}{\r(\s)}s})\sp \s=1,...,\l-1\nonumber
\ea 
\reseteqn
are given in Ref.~[\rf{us2}].  The central charge $\hat{c}(\s)=c$ of the twisted sectors is the same as the central charge of the untwisted sector.  

The stress tensors, central charges and ground state conformal weights of our interacting coset orbifolds
\(
\frac{(\frac{g}{\gs})}{\zl}
\)
are easily obtained from Eq.~(\ref{cwlr}) and the duality transformations (\ref{themap}).  The form of these results
\group{cosorbs}
\(
\hat{T}_{\sgb_{\r(\s)}/\gst}=\hat{T}_{\sgb_{\r(\s)}}-\hat{T}_{\gst}\sp
\s=1,...,\l-1
\)
\(
\hat{T}_{\sgb_{\r(\s)}}\equiv(\hat{T}_{\frac{g}{\ssz_\l}})_\s\sp
\hat{T}_{\gst}\equiv(\hat{T}_{\frac{\gs}{\ssz_\l}})_\s\sp 
\hat{T}_{\sgb_{\r(\s)}/\gst}\equiv(\hat{T}_{\frac{g/\gs}{\ssz_\l}})_\s
\label{redef}
\)
\(
\hat{c}_{\sgb_{\r(\s)}/\gst}=c_{g/\gs}=\l x\dg(\frac{1}{x+\tilde{h}}-\frac{1}{\loe x+\tilde{h}})
\)
\(
(\hat{\Delta}_0)_{\sgb_{\r(\s)}/\gst}=(\hat{\Delta}_0)_{\sgb_{\r(\s)}}-(\hat{\Delta}_0)_\gst
\)
\reseteqn
follows from the linearity of the duality transformations and the linearity of K-conjugation, and shows that these coset orbifolds can be understood in terms of the component orbifolds $g/\zl$ and $\gs/\zl$.  Here the various stress tensors of the twisted sectors are obtained by the duality transformations
\(
T_g\rightarrow\hat{T}_{\sgb_{\r(\s)}}\sp
T_{\gs}\rightarrow\hat{T}_{\gst}\sp
T_{g/\gs}\rightarrow\hat{T}_{\sgb_{\r(\s)}/\gst}
\)
of each component of $T_{g/\gs}$ in (\ref{untwistcoset}).  In the following subsections, we will consider these component orbifolds separately.

Similarly, one finds for general coset orbifolds $(g/h)/H$ that
\(
(\hat{T}_{\frac{(g/h)}{H}})_\s=(\hat{T}_{\frac{g}{H}})_\s-(\hat{T}_{\frac{h}{H}})_\s
\)
so the general case can be understood in terms of the component orbifolds $g/H$ and $h/H$.

\subsection{The WZW cyclic orbifolds $g/\zl$}
For each WZW cyclic orbifold $g/\zl$, the set of stress tensors in the twisted sectors$^{\rf{us2}}$
\group{wzwcon}
\(
(\hat{T}_{g/\sz_\l})_\s=\hat{T}_{{\sgb}_{\r(\s)}}=\frac{1}{\r(\s)}\frac{\e^{ab}}{2k+Q}\sum_{r=0}^{\r(\s)-1} \sum_{j=0}^{\lors-1}:\jh_{a(j)}^{(r)} \jh_{b(j)}^{(-r)}:\sp
\s=1,...,\l-1 \label{wzwstress}
\)
\(
\hat{c}_{\sgb_{\r(\s)}}=c_g= \frac{\l x \ \dg}{x+\tilde{h}} \sp
(\hat{\Delta}_0)_{\sgb_{\r(\s)}} = \frac{\l x \dg}{24 (x+\tilde{h})} (1-\frac{1}{\r^2(\s)}) \label{wzwcw}
\)
\reseteqn
is a list of $\s$-dependent orbifold affine-Sugawara constructions$^{\rf{9},\rf{5},\rf{us2}}$.

\subsection{The orbifolds $\gs/\zl$}\label{horb}
Substitution of the inverse inertia tensor (\ref{invarl}) into the duality transformations (\ref{themap}) yields (see App.~\ref{lameapp}) the stress tensors of the twisted sectors of the orbifolds $\gs/\zl$:
\group{gstcon}
\(
(\hat{T}_{\frac{\gs}{\ssz_\l}})_\s=\hat{T}_\gst=\sum_{r=0}^{\r(\s)-1}\sum_{j,l=0}^{\lors-1}:(\lr_r^{j-l}(\s)\hat{H}_{A(j)}^{(r)}\hat{H}_{A(l)}^{(-r)}+\lr_r^{\a,j-l}(\s)\hat{E}_{\a(j)}^{(r)}\hat{E}_{-\a(l)}^{(-r)}):
\label{gststress}
\)
\getletter{lra}
\(
\lr^{j-l}_r(\s)\equiv\frac{1}{\eta/\mu}\frac{1}{2\loe k+Q} e^{\frac{\tp (\frac{j-l}{\mu})PN(\s)r}{\r(\s)}}\delta_{j-l,0\mod\mu}\delta_{r,0\mod\frac{\r(\s)\mu}{\e}} 
\) 
\getletter{lrb}
\(
\lr^{\a,j-l}_r(\s)\equiv\frac{1}{\eta/\mu}\frac{1}{2\loe k+Q} e^{\frac{\tp (j-l)\si(\a)}{\l}}e^{\frac{\tp (\frac{j-l}{\mu})P(N(\s)r-\si(\a))}{\r(\s)}}\delta_{j-l,0\mod\mu}\delta_{N(\s)r,\si(\a)\mod\frac{\r(\s)\mu}{\e}}
\)
\(
\hat{c}_{\gst}=c_\gs=\frac{\l x \dg}{\loe x+\tilde{h}}\sp
\)
\ba
(\hat{\Delta}_0)_\gst=\frac{\l x}{24\r(\s)^2(\loe x+\tilde{h})}[&&\hsp{-.27}\dg(\r(\s)^2-1)\nonumber\\
&&\hsp{-1.266}-3\sum_{R=1}^{\frac{\r(\s)\mu}{\e}-1}csc^2(\frac{\pi R \e N(\s)}{\r(\s)\mu})(\textup{rank}\gb+\hsp{-.0666}\sum_{\a\in\Delta(\gb)}cos({\frac{2\pi R\e\si(\a)}{\r(\s)\mu}}))] \label{subcw}
\ea
\(
\a\in\Delta(\gb)\sp
A=1,...,\textup{rank}\gb\sp
j,l=0,...,\lors-1.
\)
\reseteqn
We note that the ground state conformal weights in (\ref{subcw}) depend in general on the inner-automorphic parameters $\phi(\a)\sim d\cdot\a$.

When the order $\l$ of a cyclic permutation orbifold is prime, every twisted sector $\s$ has order $\r(\s)=\l$ and so $j=l=0$.  Moreover $\mu=1$ so $R=S=0$.  As a result many of the expressions above simplify, and one obtains in particular
\alpheqn
\(
\hat{H}_A^{R=0,(r)}(m+\frac{\l r/\e}{\l})=\hat{H}_A^{(\loe r)}(m+\frac{\l r/\e}{\l})
\)
\(
\hat{E}_\a^{R=0,(r)}(m+\frac{\s\si(\a)+\loe r}{\l})=\hat{E}_\a^{(\s\si(\a)+\l r/\e)}(m+\frac{\s\si(\a)+\loe r}{\l})
\)
\(
\lr_r^{AB}(\s)=\frac{\e^{AB}\delta_{r,0\ \mod \loe}}{2\l k+\e Q}\sp
\lr_r^{\a,-\a}(\s)=\frac{\delta_{r,\s\si(\a)\ \mod \loe}}{2\l k+\e Q}
\)
\(
A,B=1,...,{\textrm{rank}}\gb\sp
\alpha,\beta\in\Delta(\gb)\sp
r,s=0,...,\e-1\sp
\loe\in\z^+
\)
\reseteqn
from (\ref{hopes}) and (\ref{gstcon}).  When $\sv=\{0\}$, these twisted $h$ currents generate the subalgebras $\gb_\e\subset\gb_\l$ of orbifold affine algebra identified in Ref.~[\rf{5}].  

\subsection{The orbifold affine-Sugawara constructions of $\gs/\zl$}

The stress tensors (\ref{gstcon}) of the twisted sectors of the orbifolds $\gs/\zl$ take the simpler form
\group{simpt}
\(
\hat{T}_\gst=\frac{\mu}{2\l k+\e Q}\sum_{R=0}^{\mu-1}\sum_{r=0}^{\frac{\e}{\mu}-1}:(\hat{H}_A^{R,(r)}(\circ)\hat{H}_A^{R,(-r)}(\circ)+\hat{E}_\a^{R,(r)}(\circ)\hat{E}_{-\a}^{R,(-r)}(\circ)):
\)
\(
\s=1,...,\l-1
\)
\reseteqn
when written in terms of the twisted $h$ currents \hc.  This form shows that the stress tensors $\hat{T}_{\gst}$ are nothing but the appropriate orbifold affine-Sugawara constructions$^{\rf{9}}$ on the twisted subalgebras $\gst$.  

To understand this, recall that $\gs$ is embedded in affine $g$ at level $\loe k$ and the OPE form\footnote{Although $\gst$ in (\ref{halg}) is not an orbifold affine algebra, it is the OPE form of an algebra that determines$^{\rf{us2}}$ the Virasoro constructions.} (\ref{realhopes}) of $\gst$ has orbifold affine order $\e/\mu$.  Then the prefactor in (\ref{simpt}) can be computed via the map
\alpheqn
\(
\textup{(level of affine $g$)}\hsp{.3}k\rightarrow\loe k\hsp{.3}\textup{(level of affine $h(\e,\Phi)$)}
\)
\(
\textup{(order of $\gr$)}\hsp{.3}\r\rightarrow\frac{\e}{\mu}\hsp{.3}\textup{(order of $\gst$)}
\)
\(
\textup{(prefactor of $\hat{T}_{\sgb_{\r(\s)}}$)}\hsp{.3}(\frac{1}{\r})(\frac{1}{2k+Q})\rightarrow(\frac{1}{\e/\mu})(\frac{1}{2\loe k+Q})\hsp{.3}\textup{(prefactor of $\hat{T}_{\gst}$)}
\)
\reseteqn
from the prefactor of the orbifold affine-Sugawara construction $\hat{T}_{\sgb_{\r(\s)}}$ in (\ref{wzwcon}).

\subsection{Twisted (0,0) operators of the interacting coset orbifolds} \label{zerosec}
The untwisted currents $J_h$ of a general coset construction $g/h$ are (0,0) operators of $T_{g/h}$, and this situation applies in the untwisted sector of the general coset orbifold
\(
\frac{(g/h)}{H}\subset\frac{A(H)}{H}\sp
h\subset g\sp H\subset Aut(g)\sp H\subset Aut(h).
\)
Then, the duality algorithm of Ref.~[\rf{us2}] tells us that the twisted $h$ currents $\hat{J}_h$ of a general coset orbifold (see Subsec.~\ref{twistsubsec}) are also twisted $(0,0)$ operators
\(
T_{g/h}(z)J_{h}(w)=\reg\ \ \rightarrow\ \ 
\hat{T}_{\frac{g/h}{H}}(z)_\s\hat{J}_{h}(w;\s)=\reg
\label{dualalg}
\)
in the twisted sectors of the general coset orbifold.  In what follows, we verify this explicitly for our class of interacting cyclic coset orbifolds.

By explicit computation$^{\rf{wick},\rf{5}}$ with the form (\ref{simpt}) and the OPE's (\ref{realhopes}), we find that
\alpheqn
\ba
\hat{T}_{\sgb_{\r(\s)}}(z)\hat{J}_a^{R,(r)}(w;\circ;\s)&=&
\hat{T}_\gst(z)\hat{J}_a^{R,(r)}(w;\circ;\s)\\&=&
(\frac{1}{(z-w)^2}+\frac{\partial_w}{z-w})\hat{J}_a^{R,(r)}(w;\circ;\s)+\reg\nonumber
\ea
\ba
[L_{\sgb_{\r(\s)}}(m),\hat{H}_{A}^{R,(r)}(n+\frac{r}{\e/\mu})]&=&
[L_{\gst}(m),\hat{H}_{A}^{R,(r)}(n+\frac{r}{\e/\mu})]\\&=&
-(n+\frac{r}{\e/\mu})\hat{H}_{A}^{R,(r)}(m+n+\frac{r}{\e/\mu})\nonumber
\ea
\ba
[L_{\sgb_{\r(\s)}}(m),\hat{E}_{\a}^{R,(r)}(n+\frac{r}{\e/\mu}+\frac{\s \si(\a)}{\l})]&=&
[L_{\gst}(m),\hat{E}_{\a}^{R,(r)}(n+\frac{r}{\e/\mu}+\frac{\s \si(\a)}{\l})]\\
&=&-(n+\frac{r}{\e/\mu}+\frac{\s \si(\a)}{\l})\hat{E}_{\a}^{R,(r)}(m+n+\frac{r}{\e/\mu}+\frac{\s \si(\a)}{\l}).\nonumber
\ea
\reseteqn
As expected, the currents of the $\gst$ subalgebra are ``twisted (1,0) operators'' of $\hat{T}_{\sgb_{\r(\s)}}$ and $\hat{T}_\gst$, and hence twisted (0,0) operators 
\alpheqn
\(
\hat{T}_{\sgb_{\r(\s)}/\gst}(z)\hat{J}_a^{R,(r)}(w;\circ;\s)=\reg
\)
\(
[L_{\sgb_{\r(\s)}/\gst}(m),\hat{H}_{A}^{R,(r)}(n+\frac{r}{\e/\mu})]=
[L_{\sgb_{\r(\s)}/\gst}(m),\hat{E}_{\a}^{R,(r)}(n+\frac{r}{\e/\mu}+\frac{\s \si(\a)}{\l})]=0
\)
\reseteqn
in each sector of the interacting cyclic coset orbifolds.

In an action formulation of coset orbifolds, one must learn to gauge these twisted $h$ subalgebras.  The twisted (0,0) operators of the cyclic coset copy orbifolds are discussed in App.~\ref{copyapp}.

\newsection{Examples}\label{exsec}
The components for $g/\zl$ in (\ref{wzwcon}) and $\gs/\zl$ in (\ref{gstcon}) and (\ref{simpt}) are now easily assembled into the stress tensors and ground state conformal weights of the twisted sectors
\alpheqn
\(
\hat{T}_{\sgb_{\r(\s)}/\gst}=\hat{T}_{\sgb_{\r(\s)}}-\hat{T}_{\gst}\sp
\s=1,...,\l-1
\)
\(
\hat{c}_{\sgb_{\r(\s)}/\gst}=c_{g/\gs}=\l x\dg(\frac{1}{x+\tilde{h}}-\frac{1}{\loe x+\tilde{h}})
\)
\(
(\hat{\Delta}_0)_{\sgb_{\r(\s)}/\gst}=(\hat{\Delta}_0)_{\sgb_{\r(\s)}}-(\hat{\Delta}_0)_\gst
\)
\reseteqn
of these interacting coset orbifolds.  The examples below use the orbifold affine-Sugawara form (\ref{simpt}) for $\hat{T}_{\gst}$ in terms of the twisted $h$ currents, while the alternate form of $\hat{T}_{\gst}$ in (\ref{gstcon}) is used in Appendix~\ref{lapp} to write the coset stress tensors entirely in terms of the inverse inertia tensors $L$ and $\lr$ and the ambient algebras $g$ and $\gb_{\r(\s)}$.

For the first three examples, we return to the general basis
\(
J_g=\{J_{a(I)}\}\sp
\hat{J}_{\sgb_{\r(\s)}}=\{\hat{J}_{a(j)}^{(r)}\};\hsp{.3}
J_{\gs}=\{J_a^R(\circ)\}\sp
\hat{J}_{\gst}=\{\hat{J}_a^{R,(r)}(\circ;\s)\}
\)
of the ambient algebras and their subalgebras.  Moreover, we will suppress the sector label $\s$ of the twisted $h$ currents.

\subsection{All $\gb,\ \l$ and $\e$ for $\sv=\{0\}$} \label{firstex}
These are the interacting coset orbifolds 
\(
\frac{\bp\hsp{-.05}{\frac{\mbox{$g$}_k}{\mbox{$h(\e,\{0\})$}}}\hsp{-.05}\cbp}{\zl}
=\frac{\BIG{(}\frac{\times_{I=0}^{\l-1}\gb_k^I}{\times_{R=0}^{\e-1}\gb_{\loe k}^R}\BIG{)}}{\zl}
\label{firstpic}
\)
where $k$ is the level of $\gb^I$.  In these cases the untwisted subalgebra $h(\e,\{0\})$ is generated by
\(
J_a^R=\sum_{j=0}^{\loe-1}J_{a(\e j+R)}\sp a=1,...,\dg\sp 
R=0,...,\e-1\sp \s=0
\)
so that $h(\e,\{0\})$ is composed of $\e$ copies, labeled by $R$, of affine $\gb$ at level $\loe k$.  The twisted subalgebra $h(\e,\{0\},\s\neq 0)$ is generated by the twisted $h$ currents
\alpheqn
\(
\hat{J}_a^{R,(r)}=\sum_{j=0}^{\frac{\l}{\r(\s)\mu}-1}e^{\frac{\tp jrPN(\s)}{\e/\mu}}\hat{J}_{a(\mu j+R)}^{(\frac{\r(\s)\mu}{\e}r)}
\)
\(
[\hat{J}_a^{R,(r)}(m+\frac{r}{\e/\mu}),\hat{J}_b^{S,(s)}(n+\frac{s}{\e/\mu})]\hsp{3}
\)
\[
\hsp{1}=\delta^{RS}\{if_{ab}^{\ \ c}\hat{J}_c^{R,(r+s)}(m+n+\frac{r+s}{\e/\mu})+\frac{\l k}{\mu}(m+\frac{r}{\e/\mu})\e_{ab}\delta_{m+n+\frac{r+s}{\e/\mu},0}\}
\]
\(
\s=1,...,\l-1\sp R,S=0,...,\mu-1\sp r,s=0,...,\frac{\e}{\mu}-1
\)
\reseteqn
where $\mu=\mu(\e,\s)$ is defined in (\ref{hopes}).  This is the algebra of $\mu$ copies of an order $\e/\mu$ orbifold affine algebra with each copy at level $\l k/\mu$.

The stress tensors, central charges and ground state conformal weights of these interacting cyclic coset orbifolds are
\alpheqn
\(
\hat{T}_{\s=0}=T_{g/h(\e,\{0\})}=\frac{\e^{ab}}{2k+Q}\sum_{J=0}^{\l-1}:J_{a(J)}J_{b(J)}:-\frac{\e^{ab}}{2\loe k+Q}\sum_{R=0}^{\e-1}:J_a^RJ_b^R:
\)
\ba
\hat{T}_{\s\neq 0}=\hat{T}_{\sgb_{\r(\s)}/h(\e,\{0\},\s)}&=&\frac{\e^{ab}}{2\r(\s)k+\r(\s)Q}\sum_{r=0}^{\r(\s)-1}\sum_{j=0}^{\lors-1}:\hat{J}_{a(j)}^{(r)}\hat{J}_{b(j)}^{(-r)}:\nonumber\\
&&-\frac{\e^{ab}}{2\frac{\l}{\mu}k+\frac{\e}{\mu}Q}\sum_{r=0}^{\frac{\e}{\mu}-1}\sum_{R=0}^{\mu-1}:\hat{J}_{a}^{R,(r)}\hat{J}_{b}^{R,(-r)}:
\ea
\(
\hat{c}(\s)=c=\l x\dg(\frac{1}{x+\tilde{h}}-\frac{1}{\loe x+\tilde{h}})
\)
\(
\hat{\Delta}_0(\s)=[\frac{\r(\s)^2-1}{\r(\s)^2(x+\tilde{h})}-\frac{\e^2-\mu^2}{\e^2(\loe x+\tilde{h})}]\frac{\l x\dg}{24}.
\)
\reseteqn
For the sectors with $\r(\s)=\l$, the stress tensors of these orbifolds were given in Ref.~[\rf{5}] where they were called $\hat{T}_{\sgb_\l/\sgb_\e}$.

\subsection{All $\gb$ and $\l$ for $\e=1$ and $\sv=\{0\}$}
To illustrate our formalism, we include a discussion of the special case  $\e=1$ of the previous example
\(
\frac{(\frac{\times_{I=0}^{\l-1}\gb_k^I}{\mbox{$\gb$}_{\l k}})}{\zl}
\)
which was also discussed in Ref.~[\rf{us2}].  The $h$ currents of these orbifolds are
\alpheqn
\(
h(1,\{0\})=\gb_{\diag}=\gb_{\l k}:\hsp{.3}
J_a^{R=0}=\sum_{J=0}^{\l-1}J_{a(J)}\equiv J_{a}^{\textup{\diag}}\sp
\s=0
\)
\(
h(1,\{0\},\s):\hsp{.3}
\hat{J}_a^{R=0,(r=0)}=\sum_{j=0}^{\lors-1}\hat{J}_{a(j)}^{(0)}\sp
\s=1,...,\l-1.
\)
\reseteqn
The untwisted $h$ currents $J_a^{R=0}$ generate the diagonal subalgebra $\gb_{\l k}$ of the ambient algebra $g$ and the twisted $h$ currents $\hat{J}_a^{R=0,(r=0)}$ generate the diagonal subalgebra of the integral affine subalgebra$^{\rf{9}}$ of the ambient orbifold affine algebra $\gb_{\r(\s)}$. 

The stress tensors, central charges and ground state conformal weights of these orbifolds are
\alpheqn
\(
\hat{T}_{\s=0}=T_{g/h(1,\{0\})}=\frac{\e^{ab}}{2k+Q}\sum_{J=0}^{\l-1}:J_{a(J)}J_{b(J)}:-\frac{1}{2\l k+Q}:J_{a}^{R=0}J_{b}^{R=0}:
\)
\ba
\hat{T}_{\s\neq 0}=\hat{T}_{\sgb_{\r(\s)}/h(1,\{0\},\s)}
&=&\frac{\e^{ab}}{2\r(\s)k+\r(\s)Q}\sum_{r=0}^{\r(\s)-1}\sum_{j=0}^{\lors-1}:\hat{J}_{a(j)}^{(r)}\hat{J}_{b(j)}^{(-r)}:\nonumber\\
&&-\frac{\e^{ab}}{2\l k+Q}:\jh_{a}^{R=0,(r=0)}\jh_{b}^{R=0,(r=0)}:
\ea
\(
\hat{c}(\s)=c=\l x\dg(\frac{1}{x+\tilde{h}}-\frac{1}{\l x+\tilde{h}})
\)
\(
\hat{\Delta}_0(\s)=\frac{\l x \dg}{24 (x+\tilde{h})} (1-\frac{1}{\r^2(\s)})\sp
\s=1,...,\l-1.
\)
\reseteqn
In this case the conformal weights are those of $\hat{T}_{\sgb_{\r(\s)}}$ in (\ref{wzwcon}) because the integral affine subalgebra $\{\hat{J}^{(0)}\}$ acts$^{\rf{9},\rf{5},\rf{us2}}$ on $|0\rangle_\s$ as an ordinary affine algebra.

In the twisted sector $\s=1$, we have 
\(
\r(\s=1)=\l\sp
\jh_{a}^{(r)}\equiv\jh_{a(0)}^{(r)}\sp
\jh_{a}^{R=0,(r=0)}=\jh_a^{(0)}
\)
so the stress tensor simplifies to
\(
\hat{T}_{\s=1}=\hat{T}_{\sgb_\l/\sgb_1}=\frac{\e^{ab}}{2\l k+\l Q}\sum_{r=0}^{\l-1}:\hat{J}_{a}^{(r)}\hat{J}_{b}^{(-r)}:-\frac{\e^{ab}}{2\l k+Q}:\jh_{a}^{(0)}\jh_{b}^{(0)}:.
\)
These are the Ka\u{c}-Wakimoto coset constructions, given in Ref.~[\rf{13}] and further studied in Ref.~[\rf{14}].

\subsection{All $\gb$ for $\l=4,\ \e=2$ and $\sv=\{0\}$}\label{thirdex}
Another special case of (\ref{firstpic}) is
\(
\frac{(\frac{\sgb_k\times\sgb_k\times\sgb_k\times\sgb_k}{\sgb_{2k}\times\sgb_{2k}})}{\z_4}. \label{thirdpic}
\)
These orbifolds have $\l=4$ sectors labeled by $\s=0,1,2$ and $3$ with $\r(0)=1,\ \r(1)=\r(3)=4$\ and $\r(2)=2$.  The ambient algebra of the untwisted sector $\s=0$ is $g=\times_{I=0}^3\gb^I$, and the generators of the untwisted $h$ subalgebra 
\(h(2,\{0\},\s=0)=h(2,\{0\})\subset g
\)
are the combinations
\(
J_a^{R=0}=J_{a(0)}+J_{a(2)}\sp
J_a^{R=1}=J_{a(1)}+J_{a(3)}.
\)
These generate the $h$ subalgebra $\gb_{2k}\times\gb_{2k}$ in (\ref{thirdpic}).

The sectors $\s=1$ and $\s=3$ both have order $\r(\s)=4$ and $\mu=1$, and $N(\s=1)=1,\ N(\s=3)=3$.  The ambient orbifold affine algebra in these sectors is $\gb_{\l=4}$ on simple $\gb$, and the twisted $h$ subalgebras are isomorphic
\alpheqn
\(
h(2,\{0\},\s=1)=h(2,\{0\},\s=3)\subset\gb_{\l=4}
\)
\(
\hat{J}_a^{R=0,(r=0)}=\hat{J}_a^{(0)}\sp
\hat{J}_a^{R=0,(r=1)}=\hat{J}_a^{(2)}.
\)
\reseteqn
The explicit form of this twisted $h$ subalgebra is
\ba
[\hat{J}_a^{R=0,(r)}(m+\frac{r}{2}),\hat{J}_b^{R=0,(s)}(n+\frac{s}{2})]&&\\
&&\hsp{-1.666}=if_{ab}^{\ \ c}\hat{J}_a^{R=0,(r+s)}(m+n+\frac{r+s}{2})+2k(m+\frac{r}{2})\e_{ab}\delta_{m+n+\frac{r+s}{2},0}\nonumber
\ea
where $r$ and $s$ can be $0$ or $1$.  This is the orbifold affine subalgebra called $\gb_{\e=2}\subset\gb_{\l=4}$ in Ref.~[\rf{5}].

Finally, the sector $\s=2$ has order $\r(\s=2)=\mu=2$ and $N(\s=2)=1$.  The ambient orbifold affine algebra is $\gb_2=\gb_2^0\times\gb_2^1$ (see (\ref{grho})), and the twisted $h$ subalgebra 
\(
h(2,\{0\},\s=2)\subset\gb_2=\gb_2^0\times\gb_2^1
\)
is generated by
\(
\hat{J}_a^{R=0,(r=0)}=\hat{J}_{a(0)}^{(0)}\sp
\hat{J}_a^{R=1,(r=0)}=\hat{J}_{a(1)}^{(0)}.
\)
This is the integral affine subalgebra$^{\rf{9}}$ of $\gb_2$.

The stress tensors, central charges and ground state conformal weights of these orbifolds are
\alpheqn
\(
\hat{T}_{\s=0}\=T_{g/h(2,\{0\})}\=\frac{\e^{ab}}{2k\+Q}\sum_{I=0}^3:J_{a(I)}J_{b(I)}:-\frac{\e^{ab}}{4k\+Q}:(J_a^{R=0}J_b^{R=0}\+J_a^{R=1}J_b^{R=1}):
\)
\ba
\hsp{-.3}\hat{T}_{\s=1}\=\hat{T}_{\s=3}\=\hat{T}_{\sgb_4/\sgb_2}&\=&\frac{\e^{ab}}{8k\+4Q}:(\hat{J}_a^{(0)}\hat{J}_b^{(0)}\+\hat{J}_a^{(1)}\hat{J}_b^{(-1)}\+\hat{J}_a^{(2)}\hat{J}_b^{(-2)}\+\hat{J}_a^{(3)}\hat{J}_b^{(-3)}):\\
&&-\frac{\e^{ab}}{8k\+2Q}:(\hat{J}_a^{R=0,(r=0)}\hat{J}_b^{R=0,(r=0)}\+\hat{J}_a^{R=0,(r=1)}\hat{J}_b^{R=0,(r=1)}):\nonumber
\ea
\ba
\hsp{-.3}\hat{T}_{\s=2}\=\hat{T}_{\sgb_2/h(2,\{0\},2)}&\=&\frac{\e^{ab}}{4k\+2Q}:(\hat{J}_{a(0)}^{(0)}\hat{J}_{b(0)}^{(0)}\+\hat{J}_{a(0)}^{(1)}\hat{J}_{b(0)}^{(-1)}\+\hat{J}_{a(1)}^{(0)}\hat{J}_{b(1)}^{(0)}\+\hat{J}_{a(1)}^{(1)}\hat{J}_{b(1)}^{(-1)}):\\
& &-\frac{\e^{ab}}{4k\+Q}:(\hat{J}_{a}^{R=0,(r=0)}\hat{J}_{b}^{R=0,(r=0)}\+\hat{J}_{a}^{R=1,(r=0)}\hat{J}_{b}^{R=1,(r=0)}):\nonumber
\ea
\(
\hat{c}(\s)=c=4x\dg(\frac{1}{x+\tilde{h}}-\frac{1}{2x+\tilde{h}})
\)
\(
\hat{\Delta}_0(\s\=1,3)\=x\dg[\frac{5}{32(x\+\tilde{h})}-\frac{1}{8(2x\+\tilde{h})}]\sp
\hat{\Delta}_0(\s\=2)\=\frac{x\dg}{8(x\+\tilde{h})}. \label{christcw}
\)
\reseteqn

\subsection{$\gb=$SU(2) for $\l=3$,\ $\e=1$ and $\{\si(\a)=1\}$} \label{fourthex}
This interacting coset orbifold
\(
\frac{(\frac{SU(2)_k\times SU(2)_k\times SU(2)_k}{SU(2)_{3k}{}\p})}{\z_3}
\)
has three sectors labeled by $\s=0,\ 1$ and $2$ with $\r(0)=1$ and $\r(1)=\r(2)=3$.  We need the integers $N(\s=1)=1$, $N(\s=2)=2$, and for each of these sectors $\mu=1$ because $\e=1$.  As noted below Eq.~(\ref{newex}), the untwisted currents of $h(\e=1,\{\si(\a)=1\})$ 
\alpheqn
\(
H^{R=0}=H_{(0)}+H_{(1)}+H_{(2)}
\)
\(
E_{\pm\a}^{R=0}=E_{\pm\a(0)}+e^{\pm\frac{\tp}{3}}E_{\pm\a(1)}+e^{\mp\frac{\tp}{3}}E_{\pm\a(2)}
\)
\reseteqn
generate SU$(2)_{3k}{}\p$ at level $3k$.  

The twisted currents of $h(\e=1,\{\si(\a)=1\},\s=1)$ are
\alpheqn
\(
\hat{H}^{R=0,(r=0)}=\hat{H}^{(0)}\sp
\hat{E}_{\pm\a}^{R=0,(r=0)}=\hat{E}_{\pm\a}^{(\pm 1)}
\)
\(
[\hat{H}^{R=0,(r=0)}(m),\hat{H}^{R=0,(r=0)}(n)]=3km\delta_{m+n,0}
\)
\(
[\hat{H}^{R=0,(r=0)}(m),\hat{E}_{\pm\a}^{R=0,(r=0)}(n\pm\frac{1}{3})]=\pm\a \hat{E}_{\pm\a}^{R=0,(r=0)}(m+n\pm\frac{1}{3})
\)
\(
[\hat{E}_{\pm\a}^{R=0,(r=0)}(m\pm\frac{1}{3}),\hat{E}_{\mp\a}^{R=0,(r=0)}(n\mp\frac{1}{3})]=\pm\a \hat{H}^{R=0,(r=0)}(m+n)+3k(m\pm\frac{1}{3})\delta_{m+n,0}
\)
\reseteqn
and the twisted currents of $h(\e=1,\{\si(\a)=1\},\s=2)$ are
\alpheqn
\(
\hat{H}^{R=0,(r=0)}=\hat{H}^{(0)}\sp
\hat{E}_{\pm\a}^{R=0,(r=0)}=\hat{E}_{\pm\a}^{(\mp 1)}
\)
\(
[\hat{H}^{R=0,(r=0)}(m),\hat{H}^{R=0,(r=0)}(n)]=3km\delta_{m+n,0}
\)
\(
[\hat{H}^{R=0,(r=0)}(m),\hat{E}_{\pm\a}^{R=0,(r=0)}(n\mp\frac{1}{3})]=\pm\a \hat{E}_{\pm\a}^{R=0,(r=0)}(m+n\pm\frac{1}{3})
\)
\(
[\hat{E}_{\pm\a}^{R=0,(r=0)}(m\mp\frac{1}{3}),\hat{E}_{\mp\a}^{R=0,(r=0)}(n\pm\frac{1}{3})]=\pm\a \hat{H}^{R=0,(r=0)}(m+n)+3k(m\mp\frac{1}{3})\delta_{m+n,0}.
\)
\reseteqn
The twisted $h$ subalgebras of sectors $\s=1$ and $2$ are inner-automorphically twisted affine Lie algebras on simple $\gb$, as are all the twisted $h$ subalgebras with $d\neq 0$ and $\e=1$.  

This gives the following stress tensors, central charges and ground state conformal weights in each sector $\s$:
\alpheqn
\ba
\hsp{-.3}\hat{T}_{\s=0}=T_{g/h(1,\{1\})}&=&\frac{1}{2k+2\a^2}\sum_{I=0}^2:(H_{(I)}H_{(I)}+E_{\a (I)}E_{-\a (I)}+E_{-\a (I)}E_{\a (I)}):\nonumber\\
&-&\frac{1}{6k+2\a^2}:(H^{R=0}H^{R=0}+E_\a^{R=0}E_{-\a}^{R=0}+E_{-\a}^{R=0}E_\a^{R=0}):
\ea
\ba
\hsp{-.3}\hat{T}_{\s=1}&=&\frac{1}{6k+6\a^2}\sum_{r=0}^2:(\hat{H}^{(r)}\hat{H}^{(-r)}+\hat{E}_{\a}^{(r)}\hat{E}_{-\a}^{(-r)}+\hat{E}_{-\a}^{(r)}\hat{E}_{\a}^{(-r)}):\\
&-&\frac{1}{6k+2\a^2}:(\hat{H}^{R=0,(r=0)}\hat{H}^{R=0,(r=0)}\+\hat{E}_\a^{R=0,(r=0)}\hat{E}_{-\a}^{R=0,(r=0)}\+\hat{E}_{-\a}^{R=0,(r=0)}\hat{E}_\a^{R=0,(r=0)}):\nonumber
\ea
\ba
\hsp{-.3}\hat{T}_{\s=2}&=&\frac{1}{6k+6\a^2}\sum_{r=0}^2:(\hat{H}^{(r)}\hat{H}^{(-r)}+\hat{E}_{\a}^{(r)}\hat{E}_{-\a}^{(-r)}+\hat{E}_{-\a}^{(r)}\hat{E}_{\a}^{(-r)}):\\
&-&\frac{1}{6k+2\a^2}:(\hat{H}^{R=0,(r=0)}\hat{H}^{R=0,(r=0)}\+\hat{E}_\a^{R=0,(r=0)}\hat{E}_{-\a}^{R=0,(r=0)}\+\hat{E}_{-\a}^{R=0,(r=0)}\hat{E}_\a^{R=0,(r=0)}):\nonumber
\ea
\(
\hat{c}(\s)=c=9x(\frac{1}{x+2}-\frac{1}{3x+2})
\)
\(
\hat{\Delta}_0(\s=1)=\hat{\Delta}_0(\s=2)=\frac{x}{9}[\frac{3}{x+2}-\frac{3}{3x+2}].
\)
\reseteqn

\subsection{$\gb=$SU(2) for $\l=4$,\ $\e=2$ and $\{\si(\a)=1\}$} \label{fifthex}
Finally, we discuss an example with a doubly-twisted $h$ subalgebra
\(
\frac{(\frac{SU(2)_k\times SU(2)_k\times SU(2)_k\times SU(2)_k}{SU(2)_{2k}{}\p\times SU(2)_{2k}{}\p})}{\z_4}. 
\)
In this case, the untwisted currents $J_{\gs}$
\alpheqn
\(
H^{R=0}=H_{(0)}+H_{(2)}\sp
H^{R=1}=H_{(1)}+H_{(3)}
\)
\(
E_{\pm\a}^{R=0}=E_{\pm\a(0)}-E_{\pm\a(2)}\sp
E_{\pm\a}^{R=1}=E_{\pm\a(1)}-E_{\pm\a(3)}
\label{signdiff}
\)
\reseteqn
generate the algebra SU(2)$_{2k}{}\p\times$SU(2)$_{2k}{}\p$, which consists of two commuting copies (labeled by $R$=1,2) of an affine SU(2) at level $2k$.  Note the sign differences in (\ref{signdiff}) relative to those of the example in Subsec.~\ref{thirdex}.  The form of the stress tensor of the untwisted sector
\ba
\hsp{-.3}\hat{T}_{\s=0}=T_{g/h(2,\{1\})}&=&\frac{1}{2k+2\a^2}\sum_{I=0}^3:(H_{(I)}H_{(I)}+E_{\a (I)}E_{-\a (I)}+E_{-\a (I)}E_{\a (I)}):\nonumber\\
&-&\frac{1}{4k+2\a^2}\sum_{R=0}^1:(H^{R}H^{R}+E_\a^{R}E_{-\a}^{R}+E_{-\a}^{R}E_\a^{R}):
\ea
is not affected by these signs.

We focus on the twisted sector $\s=1$ with ambient algebra $\gb_{\r(\s=1)=4}$ and $N(\s=1)=\mu=P=1$.  The twisted currents of $h(\eta=2,\{\si(\a)=1\},\s=1)$ are
\group{dtaex}
\(
\hat{H}^{R=0,(r=0)}=\hat{H}^{(0)}\sp
\hat{H}^{R=0,(r=1)}=\hat{H}^{(2)}
\)
\(
\hat{E}_{\pm\a}^{R=0,(r=0)}=\hat{E}_{\pm\a}^{(1)}\sp
\hat{E}_{\pm\a}^{R=0,(r=1)}=\hat{E}_{\pm\a}^{(3)}.
\)
\reseteqn
These satisfy the doubly-twisted subalgebra
\alpheqn
\(
[\hat{H}^{R=0,(r)}(m+\frac{r}{2}),\hat{H}^{R=0,(s)}(n+\frac{s}{2})]=4k(m+\frac{r}{2})\delta_{m+n+\frac{r+s}{2},0}
\)
\(
[\hat{H}^{R=0,(r)}(m+\frac{r}{2}),\hat{E}_{\pm\a}^{R=0,(s)}(n+\frac{s\pm \frac{1}{2}}{2})]=\pm\a\hat{E}_{\pm\a}^{R=0,(r+s)}(m+n+\frac{r+s\pm\frac{1}{2}}{2})
\)
\ba
[\hat{E}_{\pm\a}^{R=0,(r)}(m+\frac{r\pm \frac{1}{2}}{2}),\hat{E}_{\mp\a}^{R=0,(s)}(n+\frac{s\mp \frac{1}{2}}{2})]&&\\
&&\hsp{-1.8666}=\pm\a \hat{H}^{R=0,(r+s)}(m+n+\frac{r+s}{2})+4k(m+\frac{r\pm \frac{1}{2}}{2})\delta_{m+n+\frac{r+s}{2},0}\nonumber
\ea
\reseteqn
with order $2$ and level $4k$.  The stress tensor, central charge and ground state conformal weight of this twisted sector are
\alpheqn
\ba
\hat{T}_{\s=1}&=&\frac{1}{8k+8\a^2}\sum_{r=0}^3:(\hat{H}^{(r)}\hat{H}^{(-r)}+\hat{E}_{\a}^{(r)}\hat{E}_{-\a}^{(-r)}+\hat{E}_{-\a}^{(r)}\hat{E}_{\a}^{(-r)}):\\
&-&\frac{1}{8k+4\a^2}\sum_{r=0}^1:(\hat{H}^{R=0,(r)}\hat{H}^{R=0,(r)}\+\hat{E}_\a^{R=0,(r)}\hat{E}_{-\a}^{R=0,(r)}\+\hat{E}_{-\a}^{R=0,(r)}\hat{E}_\a^{R=0,(r)}):\nonumber
\ea
\(
\hat{c}(\s=1)=c=12x(\frac{1}{x+2}-\frac{1}{2x+2})
\)
\(
\hat{\Delta}_0(\s=1)=\frac{x}{8}(\frac{15}{4(x+2)}-\frac{2}{x+1}) \label{lastcw}.
\)
\reseteqn
Note that $\hat{\Delta}_0(\s=1)$ in (\ref{lastcw}) is not the same as $\hat{\Delta}_0(\s=1)$ in (\ref{christcw}) for $\gb=$SU(2).  The two situations differ only in that the inner-automorphic vector $d\neq 0$ for the present example.

The sector $\s=3$ contains another doubly-twisted $h$ subalgebra similar to (\ref{dtaex}) and the twisted $h$ subalgebra of sector $\s=2$ consists of two commuting copies of an inner-automorphically twisted affine SU(2) at level $2k$.
\bigskip

\noindent
{\bf Acknowledgements} 

We thank C. Schweigert for a helpful discussion.  

J. E. W. was supported by the Department of Education, GAANN.  J. E. and M. B. H. were supported in part by the Director, Office of Science, Office of High Energy and Nuclear Physics, Division of High Energy Physics, of the U.S. Department of Energy under Contract DE-AC03-76SF00098 and in part by the National Science Foundation under grant PHY95-14797.

\appendices

\app{copyapp}{(0,0) operators in cyclic coset copy orbifolds}
In this appendix, we find the twisted (0,0) operators (which are the twisted $h$ currents) in the simpler case of the cyclic coset copy orbifolds$^{\rf{39},\rf{40},\rf{9},\rf{41},\rf{5},\rf{us2},\rf{5c},\rf{5d}}$ 
\(
\times_{I=0}^{\l-1}(\gb/\hb)^I/\zl.
\)
The stress tensors in the untwisted sectors of these orbifolds are sums of $\l$ commuting copies of the coset construction $\gb/\hb$, $\hb\subset\gb$.  The Lie algebras involved here are
\(
g=\times_{I=0}^{\l-1}\gb^I\sp
\gb^I\cong\gb;\hsp{.4}
h=\times_{I=0}^{\l-1}\hb^I\sp
\hb^I\cong\hb\subset\gb
\)
and the ambient affine algebra is still given by (\ref{affalg}).  The embedding of the untwisted $h$ currents $J_{A(I)}$ in the untwisted $g$ currents $J_{a(I)}$ is
\alpheqn
\(
J_{A(I)}(z)=M_A^{\ a}J_{a(I)}(z)
\)
\(
A=1,...,\dh\sp
a=1,...,\dg\sp
I=0,...,\l-1
\)
\reseteqn
where $M_A^{\ a}$ is the embedding matrix of $\hb\subset\gb$.

In this case, the induced action $\Omega$ of $\zl$(permutation) on the untwisted $h$ currents 
\(
J_{A(I)}(z)\ \p=\w_{IJ}(h_\s)J_{A(J)}(z)=J_{A(I+\s)}(z)\sp
\Omega(h_\s)=\w(h_\s)\sp\s=0,...,\l-1
\)
is the \textit{same} as the action $\w$ in Eq.~(\ref{zlact}) on the $g$ currents.  This follows because the embedding matrix $M_A^{\ a}$ does not mix values of the index $I$.  Then the duality transformations of Ref.~[\rf{us2}] tell us that the twisted $h$ currents $\hat{J}_{A(j)}^{(r)}$ are the \textit{same} linear combinations\footnote{That is, $\hat{M}(\s)=M$ in the language of Subsec.~\ref{twistsubsec}.} of the currents of the ambient orbifold affine algebra $\gb_{\r(\s)}$
\alpheqn 
\(
\hat{J}_{A(j)}^{(r)}(z)=M_A^{\ a}\hat{J}_{a(j)}^{(r)}(z)\sp
\hat{J}_{A(j)}^{(r)}(ze^{\tp})=e^{-\frac{\tp r}{\r(\s)}}\hat{J}_{A(j)}^{(r)}(z) 
\label{tj}
\)
\(
\hb_{\r(\s)}=\times_{j=0}^{\lors-1}\hb_{\r(\s)}^j\subset\gb_{\r(\s)}.
\)
\reseteqn
Here each $\hb_{\r(\s)}^j$ is a copy of an order $\r(\s)$, level $\r(\s)k$ orbifold affine algebra on simple $\hb$.

For these orbifolds, the stress tensor of sector $\s$ is$^{\rf{us2}}$
\alpheqn
\(
\hat{T}_{\sgb_{\r(\s)}/\shb_{\r(\s)}}(z)=\hat{T}_{\sgb_{\r(\s)}}(z)-\hat{T}_{\shb_{\r(\s)}}(z)
\)
\(
\hat{T}_{\sgb_{\r(\s)}}(z)=\frac{1}{\r(\s)}\frac{\e^{ab}}{2k+Q_{\sgb}}\sum_{r=0}^{\r(\s)-1}\sum_{j=0}^{\lors-1}:\hat{J}_{a(j)}^{(r)}(z)\hat{J}_{b(j)}^{(-r)}(z):
\)
\(
\hat{T}_{\shb_{\r(\s)}}(z)=\frac{1}{\r(\s)}\frac{\e^{AB}}{2k+Q_{\shb}}\sum_{r=0}^{\r(\s)-1}\sum_{j=0}^{\lors-1}:\hat{J}_{A(j)}^{(r)}(z)\hat{J}_{B(j)}^{(-r)}(z):
\)
\reseteqn
where the untwisted sector is recovered when $\s=0$ and hence $\r(\s=0)=1$.  The coset central charge and the ground state conformal weight of each sector $\s$ are$^{\rf{us2}}$
\(
\hat{c}_{\sgb_{\r(\s)}/\shb_{\r(\s)}}=c_{g/h}
=\l x_{\sgb}(\frac{\dg}{x_{\sgb}+\tilde{h}_\sgb}-\frac{r\dh}{rx_{\sgb}+\tilde{h}_\shb})\sp
\hat{\Delta}_0(\s)=\frac{c_{g/h}}{24}(1-\frac{1}{\r(\s)^2})
\)
where $r$ is the index of embedding of $\hb$ in $\gb$.  The twisted $h$ currents $\hat{J}_{A(j)}^{(r)}$ in (\ref{tj}) are twisted (1,0) operators of both stress tensors $\hat{T}_{\sgb_{\r(\s)}}$ and $\hat{T}_{\shb_{\r(\s)}}$, and so they are twisted (0,0) operators in each sector of each cyclic coset copy orbifold.

\app{lameapp}{Induced action of $\zl$(permutation)}
To obtain the induced action (\ref{bigomega}) of the $\zl$ automorphisms on the untwisted $h$ currents $J_{\gs}$, start with Eq.~(\ref{thedefs}) and follow the steps:
\alpheqn
\(
E_\a^R\ \p=\sum_{j=0}^{\loe-1}e^{\frac{\tp j\phi(\a)}{\l/\e}}E_{\a(\e j+R+\s)}
\)
\(
R+\s=\e\lfloor\frac{R+\s}{\e}\rfloor+S\sp
S\in\{0,...,\e-1\}
\)
\(
E_\a^R\ \p=\sum_{j=0}^{\loe-1}e^{\frac{\tp j\phi(\a)}{\l/\e}}E_{\a(\e(j+\lfloor\frac{R+\s}{\e}\rfloor)+R+S)}
\)
\(
j\rightarrow j-\lfloor\frac{R+\s}{\e}\rfloor
\)
\ba
E_\a^R\ \p&=&\sum_{j=-\lfloor\frac{R+\s}{\e}\rfloor}^{\loe-1-\lfloor\frac{R+\s}{\e}\rfloor}e^{\frac{\tp j\phi(\a)}{\l/\e}}e^{-\frac{\tp \lfloor\frac{R+\s}{\e}\rfloor\phi(\a)}{\l/\e}}E_{\a(\e j+S)}\nonumber\\
&=&e^{-\frac{\tp \lfloor\frac{R+\s}{\e}\rfloor\phi(\a)}{\l/\e}}E_\a^S=e^{-\frac{\tp \lfloor\frac{R+\s}{\e}\rfloor\phi(\a)}{\l/\e}}E_\a^{R+\s}.
\ea
\reseteqn
The induced action on the Cartan generators in (\ref{bigomega}) follows similarly without the phases.  

The reader may find the identities
\alpheqn
\(
\delta_{\lor s+j-l,0\mod\e}=\delta_{j-l,0\mod\mu}\delta_{s+(\frac{j-l}{\mu})P,0\mod\frac{\e}{\mu}}
\)
\(
\sum_{s=0}^{\r-1}e^{-\frac{\tp Nrs}{\r}}\delta_{s+(\frac{j-l}{\mu})P,0\mod\frac{\e}{\mu}}=\frac{\r}{\e/\mu}\delta_{r,0\mod\frac{\r}{\e/\mu}}e^{\frac{\tp Nr(\frac{j-l}{\mu})P}{\r}}
\)
\reseteqn
useful in obtaining Eq.~(\ref{gstcon}).

\app{lapp}{Ambient-algebraic form of the stress tensors}
We collect here the stress tensors $T$ and $\hat{T}$ of the class of interacting cyclic coset orbifolds
\alpheqn
\(
T_{g/\gs}=\sum_{J,L=0}^{\l-1}L^{a(J)b(L)}:J_{a(J)}J_{b(L)}:\sp\s=0
\)
\(
\hat{T}_{\sgb_{\r(\s)}/\gst}=\sum_{r=0}^{\r(\s)-1}\sum_{j,l=0}^{\lors-1}\lr_r^{a(j)b(l)}(\s):\hat{J}_{a(j)}^{(r)}\hat{J}_{b(l)}^{(-r)}:\sp\s=1,...,\l-1
\)
\(
\arange
\)
\reseteqn
discussed in the text.  Here $L$ and $\lr$ are the inverse inertia tensors related by the duality transformations$^{\rf{us2}}$ in Eq.~(\ref{themap}), and $J_{a(I)}$, $\hat{J}_{aj}^{(r)}$ are the currents of the ambient algebras $g$ and $\gb_{\r(\s)}$ respectively.

Combining Eqs.~(\ref{ascon}) and (\ref{invarform}), one finds
\alpheqn
\(
T_{g/\gs}=\sum_{J,L=0}^{\l-1}:(L^{A(J)B(L)}H_{A(J)}H_{B(L)}+L^{\a(J)\b(L)}E_{\a(J)}E_{\b(L)}):
\)
\(
L^{A(J)B(L)}\equiv\delta^{AB}(\frac{\delta^{JL}}{2k+Q}-\frac{\delta_{J-L,0\mod\e}}{2\loe k+Q})
\)
\(
L^{\a(J)\b(L)}\equiv\delta_{\a+\b,0}(\frac{\delta^{JL}}{2k+Q}-\frac{\delta_{J-L,0\mod\e}}{2\loe k+Q}e^{\frac{\tp(J-L)\si(\a)}{\l}})
\)
\(
A,B=1,...,\rrank\gb\sp\a,\b\in\Delta(\gb)
\)
\reseteqn
for the untwisted sectors.  Similarly, combining (\ref{wzwcon}) and (\ref{gstcon}), one finds
\alpheqn
\(
\hat{T}_{\sgb_{\r(\s)}/\gst}=\sum_{r=0}^{\r(\s)-1}\sum_{j,l=0}^{\lors-1}:(\lr_r^{A(j)B(l)}(\s)\hat{H}_{A(j)}^{(r)}\hat{H}_{B(l)}^{(-r)}+\lr_r^{\a(j)\b(l)}(\s)\hat{E}_{\a(j)}^{(r)}\hat{E}_{\b(l)}^{(-r)}):
\)
\(
\lr_r^{A(j)B(l)}(\s)\equiv\delta^{AB}(\frac{1}{\r}\frac{\delta^{jl}}{2k+Q}-\frac{\delta_{j-l,0\mod\mu}}{2\l k+\e Q}\mu\hsp{.02}\delta_{r,0\mod\frac{\r\mu}{\e}}e^{\tp\chi_1})
\)
\(
\lr_r^{\a(j)\b(l)}(\s)\equiv\delta_{\a+\b,0}(\frac{1}{\r}\frac{\delta^{jl}}{2k+Q}-\frac{\delta_{j-l,0\mod\mu}}{2\l k+\e Q}\mu\hsp{.02}\delta_{Nr,\si(\a)\mod\frac{\r\mu}{\e}}e^{\tp\chi_2})
\)
\(
{}_{\chi_1\hsp{.03}=\hsp{.03}\frac{j-l}{\r\mu}NPr,}\hsp{.3}
{}_{\chi_2\hsp{.03}=\hsp{.03}(j-l)\phi(\a)[\frac{1}{\l}-\frac{P}{\r\mu}]+\frac{j-l}{\r\mu}NPr}
\)
\reseteqn
for the twisted sectors.  Here we have suppressed the $\s$ and $\e$ dependence of $\r=\r(\s)$, $N=N(\s)$, $\mu=\mu(\e,\s)$ and $P=P(\e,\s)$.

\end{document}